# Low-Energy Electron Reflectivity of Graphene on Copper and other Substrates

N. Srivastava, Qin Gao, M. Widom, and R. M. Feenstra[*]
Dept. Physics, Carnegie Mellon University, Pittsburgh, PA 15213

Shu Nie and K. F. McCarty
Sandia National Laboratories, Livermore, CA 94550

I. V. Vlassiouk
Oak Ridge National Laboratory, P.O. Box 2008, Oak Ridge, TN, USA 37831

**Abstract**

The reflectivity of low energy electrons from graphene on copper substrates is studied both experimentally and theoretically. Well-known oscillations in the reflectivity of electrons with energies 0 – 8 eV above the vacuum level are observed in the experiment. These oscillations are reproduced in theory, based on a first-principles density functional description of interlayer states forming for various thicknesses of multilayer graphene. It is demonstrated that $n$ layers of graphene produce a regular series of $n-1$ minima in the reflectance spectra, together with a possible additional minimum associated with an interlayer state forming between the graphene and the substrate. Both (111) and (001) orientations of the copper substrates are studied. Similarities in their reflectivity spectra arise from the interlayer states, whereas differences are found because of the different Cu band structures along those orientations. Results for graphene on other substrates, including Pt(111) and Ir(111), are also discussed.

## I. Introduction

The reflectivity of low-energy electrons from various thin films, for energies of typically 0 – 20 eV above the vacuum level, has been known for many decades to produce an interesting spectrum with a series of minima and maxima located at energies below the first Bragg peak (the latter arising from an inter-planar separation in the film).[1,2,3,4] This pattern in low-energy electron reflectivity (LEER) spectra has been interpreted for the case of thin metal films in terms of quantum-well-type standing waves in the films.[2,4] For the case of graphene on SiC, Hibino et al. demonstrated several years ago that LEER spectra provide a useful probe of the number of graphene layers in multilayer graphene.[5] However, that work took a different approach in interpreting the spectra, proposing that the observed minima in the reflectance could be understood essentially as tight-binding-like states of the graphene layers in the film. On this basis they argued that a film containing $n$ monolayers of graphene (where $n$ is an integer) would produce $n$ local minima in its reflectivity spectra.

Recently, it was demonstrated that for free-standing graphene, a film consisting of $n$ graphene monolayers (ML) actually produces $n-1$ minima in its LEER spectrum.[6,7] The reason that $n-1$ minima are obtained, rather than $n$, is that the wavefunctions for the relevant scattering states are localized *in between* the graphene layers, not *on* them. These states derive from the *interlayer*

---

[*] Electronic mail: feenstra@cmu.edu

*band* of graphite, the structure of which depends sensitively on the exchange-correlation potential in the material.[8,9,10]

In this work we present LEER spectra of single- and multi-layer graphene on copper substrates, and we discuss spectra obtained from other substrates as well. We compare the experimental results with theoretically predicted spectra, obtained using a first-principles method. In comparison to free-standing graphene, it is demonstrated that for an *n*-layer graphene film on a substrate an additional reflectivity minimum can form associated with an interlayer state localized in the space between the graphene and the substrate. The energy of this state depends sensitively on the separation between the graphene and the substrate. Hence, by comparison of LEER spectra determined experimentally and theoretically, one obtains information about the separation (and the potential) between the graphene and the substrate.

The precise energy position of the interlayer state forming between graphene and the substrate requires a full band structure computation and subsequent analysis to determine. However, given that energy, we find that LEER spectrum for additional layers of graphene is quite simple to predict, as illustrated in Fig. 1. We let $E_1(d)$ denote the energy of the interlayer state between the substrate and the bottom-most graphene layer, separated by a distance $d$. $E_0$ denotes the energy of the interlayer states between higher-lying graphene layers, and nearest-neighbor interlayer states are coupled together with a hopping matrix element of *t* (we assume the same *t* value for coupling between the $E_0$ and $E_1$ states as for between neighboring $E_0$ states). Values of $E_0 \approx 3 \, \text{eV}$ and $t \approx 1.5 \, \text{eV}$ were estimated by Hibino et al.,[5] although, again, we are re-interpreting those as referring to interlayer states rather than states localized on the graphene planes. The resulting spectrum of *n* states for *n*-layer graphene is then easily estimated by solving for the eigenvalues of a $n \times n$ matrix with $E_1$ in the first diagonal position, $E_0$ in all other diagonal positions, and $-t$ in all off-diagonal positions immediately above and below the diagonal. The tight-binding spectrum of states is illustrated in Fig. 1(b) for the case of $E_1 = 8 \, \text{eV}$. In a LEER spectrum, a minimum in electron reflectivity is expected at each of these energies. For a single layer of graphene on the substrate, $n = 1$, the energy of the reflectivity minimum equals the energy $E_1$ of the interlayer state. Although in the present work we employ a first-principles method for computing the full reflectivity spectrum, the results thus obtained follow the tight-binding expectation of Fig. 1 fairly closely, with the main unknown in that picture being $E_1(d)$.

The situation we consider in this work of graphene on various metallic substrates is interesting for a number of reasons. For one thing, a wide range of interface structures are found for graphene on the various substrates,[11,12,13,14,15,16,17,18, 19,20,21,22] with experimental determination of the structures being not necessarily straightforward (so that development of a new tool for structural determination is useful). Furthermore, graphene formed in this way can be transferred off of the substrate,[23] and knowledge of the precise interface structure is clearly important for understanding that process. Finally, theoretical prediction of the interface geometry is also found to be difficult,[24,25] thus underscoring the importance of experiment in these sorts of studies.

For Cu(111) and Cu(001), we find average separations between the substrate and the graphene of 3.25±0.04 and 3.61±0.07 Å, respectively. For Ir(111), comparing our computations to the experimental data of Loginova et al.,[13] we find a separation between the substrate and the graphene of 3.51±0.05 Å. For Pt(111), comparing with the data of Sutter et al.,[14] we find that the structure suggested by those authors of a simple weakly-bound graphene layer on the Pt is *not* consistent with the observed LEER spectrum. Rather, inclusion of an intercalated carbon layer between the graphene and the Pt, as proposed in the early work of Zi-pu et al.,[11] produces a computed spectrum in better agreement with the observations. In general, a full interpretation of the reflectivity spectra is shown to involve, in addition to the interlayer states, consideration of the electronic band structure of the substrate along the surface normal direction insofar as band gaps can exist in that band structure.

In the following section we discuss the experimental work for both Cu(001) and Cu(111), summarizing the methods and presenting the results. In Section III we discuss the theoretical results, first for the Cu surfaces and then for other metal surfaces. Section IV discusses in qualitative terms a comparison of the theory with experiment for additional surfaces. In Section V we summarize our work, and we also briefly compare our computational methodology with that of Krasovskii et al.,[26,27] who have successfully described LEER spectra from surfaces of two-dimensional (2D) materials such as $TiS_2$ and $NbSe_2$.

## II. Experimental

### A. Cu(001)

Graphene was prepared on thin Cu foils by the chemical-vapor deposition (CVD) technique of heating Cu to below its melting point under a flow of methane and hydrogen.[23,28] By varying the flow rate of methane, graphene domains of differing shapes, sizes and thicknesses are obtained. Two samples are described here, called #1 and #2. The foils are polycrystalline, 25 μm thick and 99.999% pure. For cleaning, the foil is annealed at 1000°C for 30 minutes under 500 sccm of 2.5% $H_2$ in Ar. For graphene growth, $CH_4$ in Ar is flowed over the foil while it is continued to be heated at 1000°C. For sample #1, 10 sccm of 0.1% $CH_4$ in Ar is used and the sample is heated for 90 minutes, while for sample #2, 20 sccm of 0.1% $CH_4$ in Ar is used for 30 min. The samples were studied in an Elmitec LEEM III instrument, with brief outgassing at 250°C performed prior to measurement.

Figure 2 shows data from sample #1. This data was acquired within a few days of preparing the sample, a condition that we refer to as "freshly prepared". The graphene crystals on the surface are seen to form hexagonal-shaped (or merged hexagons) areas. A low-energy electron diffraction (LEED) pattern from one of those areas reveals the characteristic six-fold pattern of graphene, shown in Fig. 2(c), whereas a LEED pattern acquired away from those areas reveals a square arrangement of spots associated with a Cu(001) surface, as in Fig. 2(d). A diffraction spot from Cu(001) is also seen in the pattern of Fig. 2(c) acquired from the graphene. The vast majority of surface regions that we have investigated similarly show the square LEED arrangement associated with the (001) surface. Nie et al. have shown that these surfaces actually consist of alternating (001) and (410) facets,[20] with the facets giving rise to a fine striped structure in the LEEM images. Our data shows a similar striped structure, barely visible in Fig. 2(a) but slightly clearer in Fig. 3(a) particularly if compared with the data of Nie et al., indicating the same type of surface orientation with dominant (001) structure. The areas from which LEER spectra are extracted are marked on the LEEM image of Fig. 2(a); curves A and B are from bare Cu, while C – E are from graphene.

In spectra A and B of Fig. 2(b), the main feature is a sharp onset as a function of decreasing energy occurring at about 2.5 eV. The energy scale of our LEER spectra is simply the voltage difference between LEEM electron emitter and sample, multiplied by the fundamental charge. The onset near 2.5 eV corresponds to the voltage for which the vacuum levels near the sample and near the electron emitter are aligned, i.e. zero net field occurs in the space between them, as discussed by previous authors.[29] Thus, the onset energy provides a measure of the vacuum level position for each particular spectrum. The sample work function is typically higher than that of the $LaB_6$ thermionic electron emitter in the LEEM, so that the LEER spectra generally have an onset that occurs at positive energy. Aside from this vacuum level onset, the reflectivity curves from the bare Cu are nearly featureless, although they do display a slightly increased reflectivity over the energy range 2.5 – 5 eV compared to higher energies. For the case of spectra C – E, acquired from the graphene-covered areas, they display distinct minima in the reflectivity near 5.3 eV and 1.9 eV, with the vacuum level onset occurring just below the 1.9 eV minimum.

Figure 3 shows another data set from the same sample, again freshly prepared. In this case, two different types of Cu surface regions are apparent, accounting for the dark horizontal band extending across the image. The two types of Cu surfaces in the field of view likely have different orientations. Data from the lower portion and the extreme upper portion of the image closely resemble that of Fig. 2, whereas data from the central part of the image are different. Reflectivity curves D and E are acquired from the former area, again closely resembling those of Fig. 2(b), whereas the curves A – C are quite different. Curves B and C in particular, acquired from the graphene, do not display a minimum in the reflectivity near 5.3 eV. For illustrative purposes the reflectivity data in Fig. 3 has been plotted in two different ways: Fig. 3(b) shows a plot in which the curves have been shifted arbitrarily from each other for ease of viewing [the same presentation as in Fig. 2(b)], whereas in Fig. 3(c) this shifting is not done, and furthermore, the data is shown over a wider energy range.

All of the reflectivity curves of Fig. 3(c) rise to a constant value at low energies, corresponding to the reflectivity value of unity. The position of that onset for each spectrum is determined by extrapolating the linear portion of the onset until its intersection with a line at unity reflectivity, as illustrated for curve A in the inset. The resulting vacuum level locations are shown by the small vertical lines positioned above the unity reflectivity level. The position of the vacuum energy is found to be ≈0.5 eV higher energy on the bare Cu surface than on the nearby graphene-covered surface, indicating a higher work function for the former surface. Aside from these onsets associated with the vacuum level, the other major feature apparent in the spectra of Fig. 3 is the distinct minimum near 5.3 eV for reflectivity curve E. The reflectivity for that curve rises to ≈0.7 for energies between about 1.9 and 4.2 eV. A smaller increase over the same energy range is also apparent in curve D. We argue in Section III that this increase in reflectivity arises from the presence of a band gap in the band structure along the (001) direction for Cu. The fact that the reflectivity rises to a relatively large value of ≈0.7 on the graphene-covered Cu but only to ≈0.3 on the bare Cu surface indicates a greater degree of oxidation of the latter. Indeed, we have performed scanning Auger microscopy (not shown) on similar samples and we find a much greater oxygen concentration between the graphene grains than underneath them.

The most important feature in the LEER spectra that we focus upon in our comparison with theory is the location of the reflectivity minimum, at 5.3±0.1 eV for spectrum E of Fig. 3; we refer to the experimental locations of such minima as $E_1^{expt}$. We reference these values to the vacuum level position measured in the same spectra, $E_{VAC}^{expt}$, which is 1.35±0.15 eV for spectrum E of Fig. 3. Combining errors in quadrature (square root of sum of squares) we arrive at a value of $E_1^{expt} - E_{VAC}^{expt} = 3.95 \pm 0.18$ eV, or 4.0±0.2 eV in round figures, as listed in the first column of Table I. Subsequent columns of Table I list measured values from photoemission for the onset of a nearly-free-electron (NFE) band in the substrate[30,31,32,33] (as further discussed in Section III), our measured values for the work function difference $\Delta\Phi$ between the bare metal surface (an oxidized Cu surface in the present case) and the graphene-covered surface, measured values for the work function of the bare metal surface $\Phi_M$ from the literature,[34,35,36,37] and the difference $\Phi_M - \Delta\Phi$ which corresponds to an experimental value for the work function of graphene on the metal surface.

Figures 4(a) and (b) show results for sample #1 after it sat in air for several months, a condition we refer to as "air aged", and Figs. 4(c) and (d) show results for sample #2 with a similar degree of air aging. After the air exposure, the spectra acquired from the graphene region become relatively featureless, as in curves C and D of Fig. 4(b) and curve B of Fig. 4(d). Similarly the spectra from the bare Cu surface, curves A and B of Fig. 4(b) and curve A of Fig. 4(d), are also featureless. Curve E of Fig. 4(b) displays a distinct minimum in the reflectivity at about 3.8 eV; this type of spectrum arises from two layers of graphene on the surface (as identified by prior authors and also as consistent with the analysis of Section III).[20] Sample #2 had thicker graphene coverage than sample #1, so the regions displaying spectra with the single minimum near 3.8 eV are more predominant than for sample #1, as seen by curves C and D of Fig. 4(d). These 2-ML graphene regions form "dark areas" near the center of the larger graphene "bright" 1-ML areas, the same as described previously by Nie et al.[20] Additionally, some areas of sample #2 reveal spectra such as curve E containing two distinct minima in the reflectivity, at 2.5 and 5.3 eV (the latter energy being only coincidentally the same as for curve E of Fig. 3), and we associate those spectra with the presence of three graphene layers.

Summarizing the results for reflectivity of *multilayer* graphene on the copper foils, for the 2-ML spectra of Fig. 4 we find a single well-defined reflectivity minimum near 3.8 eV, and for 3-ML graphene there are two minima near 2.5 and 5.3 eV. These results are relatively independent of the underlying Cu surface (i.e. oxidized or non-oxidized), and they are consistent with expectations based on a model of a free-standing graphene film, with $n-1$ minima in the reflectivity expected for an *n*-monolayer thick film, as discussed in prior work.[6,7] More complex behavior occurs for a *single* ML of graphene on the surface, where we observe different spectra depending on both the surface orientation (as in Fig. 3) and the degree of surface oxidation (i.e. comparing Fig. 4 to Figs. 2 and 3).

**B. Cu(111)**

For comparison with the results obtained from Cu foils for which the dominant orientation observed is close to (001), we also present results from a single crystal Cu(111) surface. Details of the surface preparation and the graphene growth have been previously reported.[18] Briefly, the particular sample discussed here (sample #3) was prepared by evaporating carbon on the Cu(111) surface at about ~975°C. Growth was performed in an Elmitec LEEM III instrument. The sample was then cooled to room temperature and characterized by LEED and electron reflectivity, all under ultra-high vacuum conditions.

Figure 5 compares the LEER spectra of a single graphene layer (curve B) and bare Cu (curve A). The graphene-covered region has a shallow minimum in reflectivity at about 8 eV. In contrast, the reflectivity of the Cu decreases continuously until about 15 eV. The analyzed graphene island is a single crystal whose lattice is aligned in-plane to within a few degrees with the Cu lattice, as revealed by LEED (not shown).

Comparison of Figs. 3 and 5 clearly reveals that the LEER spectra from single-layer graphene on the Cu(111) surface are quite different from those on the Cu(001) surface. In the following Section we develop a theoretical approach for simulating LEER spectra in an effort to gain some understanding of how the spectra are affected by the substrate orientation and the structure of the graphene-substrate interface.

## III. Theoretical

### A. Cu(111) and Cu(001)

Details of our theoretical method are provided in the Appendix. In essence, we perform two computations of electronic band structure, one for a bulk substrate of the appropriate orientation and the other for a slab of substrate material on which there is 0, 1, or more layers of graphene (either on both sides or just one side of the slab). All of our computation are performed using the Vienna Ab-Initio Simulation Package (VASP), employing the projector-augmented wave method and the Perdew-Burke-Ernzerhof generalized gradient approximation (PBE-GGA) to the exchange-correlation functional,[38,39,40,41] with a plane-wave energy cutoff of 500 eV. Lattice parameters for the metals are taken from experiment,[42] and the graphene is lattice matched to the metal. We consider various commensurate fits of the graphene on the metal substrate, but we find only small changes in the resulting LEER spectra, as further discussed below and described in detail in the Appendix.

We align the energies of the bulk and slab computations by precisely matching the minimum in the potential at a point in the center of the slab. In the slab computation, there is >10 Å of vacuum on either side of the slab, and by examining the potential far out in this vacuum region we determine the energy of the vacuum level relative to the slab electronic states (this energy difference is found to have some dependence on the orientation of the surface, the graphene coverage on the substrate, and the graphene-substrate separation). At a given energy above the vacuum level we form a linear combination of the states of the slab that, in terms of its $(G_x, G_y) = (0,0)$ Fourier component, is exactly matched to a $+k_z$ eigenstate of the bulk computation (that bulk eigenstate is propagating completely in the $+z$ direction, i.e. into the substrate). As demonstrated in the Appendix, it turns out that with this matching we also succeed in closely matching the higher order Fourier component of the states. With these appropriately constructed linear combinations of slab states, the reflectivity can be directly evaluated.

For the case of simple metallic substrates such as Cu, the band structure in the relevant energy region is quite simple. There is only a *single* band of states in the substrate that couple to the incident electrons. This band, with onset typically at 4 eV above the Fermi energy ($E_F$), has dispersive, nearly-free-electron (NFE) character, as previously described.[34,43] Computed band structures for Cu in the (111) and (001) directions are shown in Figs. 6(a) and 7(a), respectively. For the (111) direction, the onset of the NFE band lies *below* the vacuum level, so that no direct signature of this band onset will occur in a LEER spectrum (since such spectra deal only with energies *above* the vacuum level). The vacuum level shown in Fig. 6(a) is obtained by performing a computation for a symmetric 3-layer Cu(111) slab with graphene on both sides, and then precisely aligning that to the bulk computation as described in the preceding paragraph. For the (001) direction, Fig. 7(a), multiple bands are present in the bulk band structure over the energy range of 0 – 15 eV relative to the vacuum level. However, as further discussed in the Appendix, we find that only a single band couples significantly to the slab states that, in turn, are coupled to the incident planes waves in the vacuum. Having only a *single channel* of bulk states to deal with permits a significant simplification of the analysis, although in general *multiple channels* of bulk states (each weighted by the overlap of the respective slab and bulk wavefunctions) must be included.

Computed reflectivity spectra for the (111) orientation are shown in Fig. 8. Although we have attempted a total energy computation using VASP to determine the graphene-Cu separation, it is difficult to confidently determine that parameter due to the omission of van der Waals interactions from our density functional. We therefore adopt as a starting value the 3.58 Å separation determined by the detailed analysis of Vanin et al.[24] Figure 8 shows results for 0, 1, and 2 layers of graphene on the Cu(111) surface. For 0 layers, a relatively flat spectrum is found, for 1 layer we see a shallow reflectivity minimum near 4.4 eV whereas for 2 layers we obtain a clear minimum at about 2.0 eV.

If we vary the graphene-Cu separation, the location of the reflectivity minima will vary in accordance with the model described in connection with Fig. 1. For the 1-ML case, the reflectivity minimum occurs at $E_1$, the energy of the interlayer state between the graphene and the Cu, i.e. 4.4 eV for a graphene-Cu separation of 3.58 Å. This energy depends sensitively on the separation *d*, increasing as *d* decreases. We investigate this reflectivity minimum in greater detail in Fig. 9, displaying results over a wider range of energies and for additional graphene-Cu separations. For a smaller graphene-Cu separation of 3.08 Å, this minimum shifts up in energy to about 9.3 eV, and for a separation of 2.58 Å it disappears. For a separation of 4.08 Å the minimum narrows and shifts to 2.2 eV, and at 4.58 Å it further shifts to 1.0 eV and a second minimum forms at 11.2 eV. The origin of these minima in the reflectivity is the same as for free-standing graphene slabs – the existence of an *interlayer state* (between the graphene and the substrate in the present case),[6] and the formation of a *transmission resonance* for incident electrons with energy equal to that of the interlayer state.[7]

Let us return to Fig. 8 to discuss the reflectivity spectra for 2 ML of graphene. We now expect, in general, two minima in the reflectivity arising from the (coupled) interlayer states that exist between the bottom-most graphene layer and the Cu and between the two graphene layers. For the $d = 4.08$ Å case in Fig. 8 these two minima are clearly seen. For $d = 3.58$ Å the second minimum is seen only faintly as the broad minimum near 5 eV (i.e. nearly the same location as the minimum for the same *d* value in the 1-ML situation). For $d = 3.08$ Å the second minimum occurs above 8 eV (again, as seen in the 1-ML situation), so it makes essentially no contribution to the 2-ML spectrum of Fig. 8. The correspondence for 2 ML of graphene between the results of the full computations of Fig. 8 and the tight-binding model of Fig. 1 is illustrated in Fig. 10. Further varying the graphene-Cu separation to much smaller or much larger values than those shown in Fig. 8 will yield just a single reflectivity minimum in the 2-ML spectrum, at an energy approaching ≈2.7 eV (or 3 eV for the illustrative example of Fig. 10). That minimum is the same as the one at 2.7 eV found for a free-standing graphene bilayer,[6] associated with the interlayer state localized between the two graphene layers. For the 2 ML of graphene on the Cu substrate, if the graphene-substrate separation is either very small or very large then the energy of the interlayer state between the bottom-most graphene layer and the Cu substrate will be too high or too low, respectively, to make a significant contribution to the spectrum. For computations of thicker multilayer graphene on Cu (not shown), we find additional well-defined minima centered at 2.7 eV, i.e. the same result as illustrated in Fig. 1.

In order to determine the actual value of the graphene-Cu separation in the experiment, we compare the computed values for the location of the low-energy reflectivity minimum to the value from experiment. In our computations we encounter the well-known problem of the predicted energies of the unoccupied states being too low.[44] The second column of Table II lists our computed energies for the onset of the NFE band, for various metals and surface orientations, which can be compared to the known values based on photoemission as listed in the second column of Table I.[30,31,32,33] In order to compare our computed reflectivity results to experiment, we shift the energy scale of the computed reflectivity up by this energy difference, i.e. $\delta E_{NFE} = 0.45 \pm 0.1$ eV for Cu(111).

From our computed LEER spectra we deduce the energy of the low-energy reflectivity minimum relative to the computed vacuum level, $E_1^{comp} - E_{VAC}^{comp}$. In principle a correction would be needed on this energy to account for the strain in the graphene for the structures we employ, e.g. a 3.5% expansion for a 1×1 arrangement of graphene on Cu(111), compared to the experimental situation in which very little strain occurs.[21] Furthermore, the energy of the reflectivity minimum might also depend on the rotational orientation and registry of the graphene on the metal surface. As detailed in the Appendix, we find that the combined effect of strain, orientation, and registry, as well as the accuracy of the computations, produce an uncertainty of ±0.2 eV on our computed values for the reflectivity minima, but with no further correction needed on those values. We thus write our final, theoretical value for the energy of the reflectivity minimum as

$$E_1^{th} - E_{VAC}^{th} = E_1^{comp} + \delta E_{NFE} - E_{VAC}^{comp}. \quad (1)$$

In comparing the resulting values to the experimentally determined $E_1^{expt} - E_{VAC}^{expt}$ values, we would in principle have to consider an additional correction to account for a difference between computed and experimental vacuum levels (work functions) for each particular surface being considered. However, we argue below that this difference is near zero, within an uncertainty of ±0.3 eV.

The location of the reflectivity minimum in the experimental data for single-layer graphene on Cu(111), curve B of Fig. 5, is found to be at 7.9±0.2 eV relative to the vacuum energy, as listed in the first column of Table I. We compare that value to theory (interpolating between the predicted locations of the reflectivity minimum as in Fig. 11) and arrive at a graphene-Cu separation of 3.25 Å. The errors in the reflectivity minimum location are combined in quadrature with those discussed above for the corrected band onset position (±0.1 eV), strain/orientation/registry/accuracy (±0.2 eV), and the vacuum level position (±0.3 eV), yielding an overall error of ±0.42 eV. The dependence of the computed reflectivity minimum location on the graphene-Cu separation is relatively steep, as seen in Fig. 11, so that this error on the energy corresponds to an error on the separation of only ±0.04 Å. We thus find a separation of 3.25±0.04 Å between the graphene and the Cu(111) surface, as listed in the third column of Table II. This value is in excellent agreement with a recent theoretical result of 3.25 Å, obtained using exact exchange and a correlation energy from the random phase approximation.[25]

With this separation value, we go back to evaluate the errors in our computed vacuum level positions, which we have already taken above as ±0.3 eV. Our computed work function (difference between vacuum level and Fermi energy) for graphene on Cu(111) is 3.95 eV, listed in the final column of Table II. We compare that value to experiment, as listed in the final column of Table I where we take the difference between the measured bare metal work functions[34,35,36,37] and our values for the difference between the bare metal vacuum level and the vacuum level for graphene on the metal. Comparing the values in the final columns of Tables I and II, we find good agreement for the case of Cu(111) and Ir(111) surfaces, whereas discrepancies are apparent for the other surfaces. There are many possible sources for these discrepancies: (i) computed work functions from VASP, using GGA, typically underestimate work functions for clean metal surfaces by about 0.3 eV,[45,46] although the corresponding errors for graphene on metal surfaces are not presently known; (ii) for the case of the oxidized Cu(001) surface in Table I the $\Phi_M$ value is an estimate, formed by combining values from Refs. [34] and [36]; (iii) the $\Delta\Phi$ values in Table I rely on vacuum onsets measured at different points in a LEEM image, and up to a few tenths of an eV error can occur in this difference due to details of the LEEM lens alignment. Unlike the situation for the onsets of the NFE band (second columns of Tables I and II), where the prior experimental values are quite well known and we could simply make a correction to our theoretical values, the sources(s) of the discrepancies between the work function values are less obvious, and hence we do not attempt any sort of correction to account for them. Rather, we incorporate these discrepancies into an error value of ±0.3 eV for our computed vacuum level energies.

Moving to the Cu(001) substrate, the onset of the NFE band now occurs *above* the graphene-on-Cu vacuum level. Computed spectra are shown in Fig. 12 for various values of the graphene-Cu separation. We see, as expected, the rise of the reflectivity to unity as the energy decreases below the onset of the NFE band [the onset of the NFE band in Fig. 12 is relative to the vacuum level for graphene on Cu(001), so referring to Table II, we compute an onset energy of $7.15 - 4.61 = 2.54$ eV]. Comparing to the experiments of Figs. 2 and 3 for graphene on Cu(001), we found that the reflectivity as a function of decreasing energy starting from about 5.3 eV rose to a value of ≈0.7 at about 4.2 eV. As the energy is reduced further, the reflectivity is relatively constant until about 1.9 eV, after which it rises and reaches unity at 1.3 eV. This 1.3-eV onset corresponds to the vacuum level of the sample. We associate the higher-energy shoulder at 4.2 eV with the onset of the NFE band in the (001) direction, at $4.2 - 1.3 = 2.9$ eV relative to the vacuum level. This value of 2.9 eV is in reasonable agreement[47] with the result we obtain by taking the difference between the band edge position measured in prior work, 7.9 eV relative to the Fermi energy as listed in Table I,[30] and our computed work function for graphene-covered Cu(001) of 4.61 eV.

Comparing the reflectivity curves of Fig. 12 with experimental results for graphene on Cu(001), the energy of the observed reflectivity minimum [curve E of Fig. 3(c)] is 4.0±0.2 eV as listed in the first column of Table I. Interpolating between the theoretical results, as shown in the inset of Fig. 12, we find best agreement between experiment and theory for a separation of 3.61±0.07 Å.

Upon oxidation of the Cu surface, the plateau in reflectivity between about 1.9 and 4.2 eV (which we have attributed to a gap in the band structure in this direction) disappears, as is apparent by comparing curves C – E of Fig. 2 or curve E of Fig. 3 with curves C – D of Fig. 4(b) or curve B of Fig. 4(d). Those data apply to a graphene-covered surface, which has a much lower oxidation rate than a bare surface, as already discussed in Section II(A). Indeed, reflectivity of the freshly-prepared *bare* surface, curves A – B of Fig. 2 and curve D of Fig. 3, also show a much reduced reflectivity for this plateau. It is thus clear that the reflectivity is rather dependent to the condition of the surface. In a similar vein, the shapes and energies of the features in the reflectivity curves of Nie et al. for graphene on Cu(001)[20] agree well with those of Figs. 2 or 3, but the magnitude of the reflectivity in the plateau at 1.9 – 4.2 eV is only ≈0.25 (compared to 0.7 for Figs. 2 and 3). We believe that this decreased magnitude also arises from some oxidation of the sample studied in Ref. [20], since it was exposed to air for two weeks between synthesis and analysis.

## B. Other Metal Surfaces

It is interesting to compare the results for Cu described above with those for other metal substrates, and for this purpose we have made computations for graphene on Ir(111) and Pt(111). The bulk band structures for these situations are similar to that of Cu(111) except that there are non-dispersive bands in the empty states at $E - E_F \approx 15$ eV for Ir and Pt that complicate the analysis, necessitating a "multi-channel analysis" as discussed in the previous Section. For simplicity, we restrict our analysis to 0 – 9 eV relative to the vacuum level, for which a single-channel analysis suffices. Figure 13 shows results for a single layer of graphene on Ir and Pt, with a graphene-substrate separation of 3.58 Å. Both of the metals show a pronounced reflectivity minimum, at nearly the same energy, although the location of those minima is ≈ 1.3 eV higher than for Cu(111). In general, the location of the minimum depends primarily on the depth of the effective (single-particle) potential in the space between the substrate and the graphene layer, e.g. as seen by the large energy lowering of the interlayer states when alkali metals are intercalated between graphene sheets.[48] However, higher-order Fourier components of the potential will also make a contribution, since these affect the location of the onset of the dispersive NFE band. The band onset for Pt (111) and Ir(111) are listed in Table I as 7.6 and 6.8 eV relative to their Fermi energies, respectively. These are considerably higher than the 4.2 eV for Cu(111), and hence the locations of the reflectivity minima are higher for Pt and Ir than for Cu.

Comparing the results displayed in Fig. 13 to experiment, we find good agreement for the case of Ir(111). Experimental results of Loginova et al.[13] reveal a vacuum level onset for graphene on Ir(111) at electron energy of 1.3 eV, with the vacuum level for the bare Ir(111) surface occurring about 0.8 eV above that. With respect to the former, a distinct reflectivity minimum occurs at 6.5 eV, and a plateau with relatively high reflectivity values of ≈0.7 is seen extending ≈2.5 eV above the vacuum level. We associate this plateau with the band gap that exists below the onset of the NFE band, known from experiment to occur at about 2.5 eV (second column of Table I minus the final column) and seen in our theory of Fig. 13 at 2.2 eV for an average graphene-Ir separation of 3.58 Å. For the reflectivity minimum, we find in Fig. 13 a position of 5.6 eV for $d = 3.58$ Å. We find that a separation of 3.51±0.05 Å produces a position for the minimum that agrees best with the experiment, as summarized in Fig. 11. Our experimental value is in reasonable agreement with the mean value of 3.38±0.04 Å measured by the x-ray standing wave technique.[17]

It is important to note that graphene on the Ir(111) surface is corrugated, with corrugation amplitude that depends on the rotational alignment of the graphene and Ir(111) lattices.[13,17] The analysis above concerns the so-called R30 variant, which has a small corrugation amplitude of about 0.04 Å. In contrast, the R0 variant has a much larger amplitude, with peak-to-peak values reported to be 0.3 Å based on scanning tunneling microscopy,[13] 0.4 – 0.6 Å (averaged over the surface) based on x-ray standing waves analysis,[17] and 0.35 Å based on atomic force microscopy.[19] The latter value is the most direct, and likely the most accurate. It is interesting to consider what effect this corrugation will have on the energy of the reflectivity minimum. Experimentally, the location of the minimum for the R0 variant is about 0.5 eV higher than for the R30 variant,[13] with this difference possibly indicative of an actual difference in average height between the graphene layers, or perhaps arising as a result of the R0 corrugation.

To estimate the effect of the corrugation on the electron reflectivity we have undertaken a series of computations as described in the Appendix. Employing a unit cell for the graphene of 3×3, we find that the effect of the corrugation is small, less than a ±0.05 eV shift in location of the reflectivity minimum for peak-to-peak amplitudes as large as 0.6 Å. On this basis, the influence of the corrugation on reflectivity spectra for both the R0 and R30 variants is expected to be negligible. Hence we would conclude that the 0.5 eV shift observed experimentally[13] would correspond to a decrease in average graphene-substrate separation for R0 compared to R30 of 0.05 Å, from the Ir(111) curve of Fig. 11. However, the 3×3 unit cell we employ in the computations is much smaller than occur in experiment; as such they provide an estimate for the effect of short-wavelength corrugation. For much longer wavelength corrugation, we can estimate its effect simply by considering the change in minimum location $\delta E_1$ arising from some modulation $\delta d$ in the graphene-substrate separation, again from Fig. 11. For a variation, say, of $\delta d = \pm 0.175$ Å (the experimental result from Ref. [19]), the resulting shift in $E_1$ would be $+1.9/-1.3$ eV, corresponding to an upwards shift in the mean position of 0.3 eV. This result differs from the 0.5 eV shift observed experimentally by 0.2 eV, and from that value we would surmise from Fig. 11 a decrease in the R0 graphene-substrate separation, compared to R30, of only 0.02 Å. Combining the short-wavelength and long-wavelength estimates, we arrive at a value of 3.48±0.07 for the average graphene-substrate separation of the R0 variant. For the other observed rotational variants, the locations of their reflectivity minima are the same as for R30 and their corrugation amplitudes are small,[13] so we conclude that their graphene-substrate separations are essentially the same as for R30.

For the case of Pt(111), our predictions of Fig. 13 for a single weakly-bound layer of graphene on the surface do *not* compare favorably with experiment. Sutter et al.[14] have measured LEER spectra for a single layer of graphene on Pt(111), and they do not find any minimum at all in the spectrum over the energy range 0 – 9 eV; rather they see just a gradually, monotonically decreasing reflectivity over that range. If we assume a rather small separation between the graphene and the Pt, e.g. 2.58 Å, then our predicted spectrum would indeed not have any minimum (similar to the Cu case shown in Fig. 9). However, on the basis of both experiment and theory it is expected that a single layer of graphene will be relatively weakly bound to a bare Pt(111) surface,[14] and thus our predicted spectra for separations of about 3.0 Å or greater are inconsistent with the experiment.

As a possible solution for this discrepancy we consider the alternate structural model proposed in early work of Zi-pu et al.,[11] in which a layer of carbon atoms is intercalated between the graphene and the Pt. We denote this model by Pt-C-Gr; we consider two possible registries of the intercalants relative to the Pt and graphene atoms as shown in Figs. 13(b) and 13(c). Both structures have one intercalant per Pt(111)-1×1 surface unit cell, i.e. half the number of intercalants as the number of atoms in the graphene. Utilizing the structural parameters of Zi-pu et al. (1.25 Å between Pt and the intercalated C and 2.45 Å between the intercalated C and the graphene) we obtain the spectra shown in Fig. 13(a). Both spectra display no region of low reflectivity whatsoever in the range 0 – 9 eV (since the separation between the intercalated C layer and the graphene is relatively small) and in this respect they are in agreement with the observed spectrum.[14] (Other models with different intercalants, e.g. Pt, would likely yield similar LEER spectra). This model including the intercalated C layer is also consistent with higher-energy LEED I(V) features, at least according to Ref. [11], although it should be noted that the LEED results differ somewhat between Refs. [11] and [14], so the structures of the respective surfaces may be different.[14]

This model of an intercalated carbon layer between the graphene and the Pt surface might seem to be inconsistent with the observation by angle-resolved photoemission spectroscopy (ARPES) of a clear Dirac cone for the valence band (VB) structure of the system.[14] To examine this, we have computed the band structure, shown in Fig. 14 for the case of the structure pictured in Fig. 13(b). Dirac cones around the K-point are clearly present, although a band gap of 0.32 eV has opened because of the inequivalence between the two graphene atoms arising from the underlying Pt lattice. The upper Dirac cone has weight concentrated around the graphene atoms located directly above the Pt atoms in the 1st Pt layer (C $2p_z$ orbitals hybridized with Pt $6s$). The lower Dirac cone has weight concentrated around the other graphene atoms, with significant hybridization of the $2p_z$ orbital of that C atom occurring with $5d_{xz}$ and $5d_{yz}$ orbitals of the surrounding Pt atoms. For the structure pictured in Fig. 13(c), with intercalants directly below graphene atoms, the Dirac cones also survive but now the band gap is larger, 0.55 eV. Finally, for a simple graphene layer located 3.58 Å above the Pt, without intercalants, clear Dirac cones also occur and the band gap is quite small, 0.07 eV. We believe that the ARPES data of Ref. [14] does not exclude any of these models. Nevertheless, it is likely that other sorts of data can be used to verify, or exclude, the Pt-C-Gr model compared to a simple Pt-Gr structure. A complete analysis of this structural model is beyond the scope of the present work since it would have to include an incommensurate graphene layer as well as possible variation in the density and locations of intercalants, as discussed by Zi-pu et al.[11] Additionally, other models would have to be considered, such as the one proposed by Otero et al.[15] with a graphene layer covalently bonded to a Pt(111) surface that contains an ordered vacancy network. Nevertheless, based on the results of Figs. 13 and 14 we can conclude the reported LEER spectra of Ref. [14] is inconsistent with a simple weakly-bound graphene layer on Pt(111), and the model suggested by Zi-pu et al. appears to provide a good candidate for resolving this discrepancy.

## IV. Discussion

In this Section we consider the experimental LEER spectra for several additional surfaces; we have not made explicit computations for these surfaces, but nevertheless we can draw some qualitative conclusions based on the experience gained from the detailed analysis of the prior Section. Considering first the case of graphene on Ni(111), for a simple, non-reconstructed Ni surface we expect similar LEER spectra as shown in Figs. 8 and 9. Experimentally for the Ni(111) surface, two layers of graphene display a distinct reflectivity minimum at $\approx 3$ eV,[22] as expected for an interlayer state between those two layers. However, the experimental result for a single graphene layer on that surface displays a relatively featureless reflectivity with roughly constant magnitude over energies of 1 – 15 eV (this magnitude is substantial, $\approx 0.7$ of the relatively large reflectivity at 12 eV for the 2-ML case).[22] That result is qualitatively in agreement with the theory of Fig. 9 if we assume a relatively small separation of $\leq 2.5$ Å between the Ni and the graphene. That is, the graphene is strongly bonded to the Ni, consistent with prior results and expectations for this system.[14,22]

Similarly, for graphene on Ru(0001), we again expect LEER spectra similar to those of Figs. 8 and 9 if the Ru surface has a simple, non-reconstructed structure. Experimentally the situation is somewhat similar to the Ni(111) surface, with a distinct reflectivity minimum at $\approx 4.5$ eV for two layers of graphene,[12] as expected from our theory (allowing for an upward shift of 1 – 2 eV in the minimum location due to the effects of the work function and/or LEEM alignment). This minimum in the LEER spectrum was successfully modeled by Sutter et al., although their modeling also produced a minimum at 11 eV that is not seen in the experiment. In any case, the LEED I(V) theory used for that modeling is of the older, "traditional" type, for which the electronic structure of the solid is not well described, particularly at low energies.[49,50,51,52] For single-layer graphene, the reflectivity is observed to be relatively constant over 3 – 10 eV,[12] although below about 3 eV a sharp drop occurs that is presently not understood. This relatively constant reflectivity, interpreted qualitatively on the basis of Figs. 8, 9, and 13, would indicate either a strongly bound single graphene layer on the surface or the presence of an intercalated layer between the graphene and the Ru, analogous to the Pt case above. This expectation is consistent with the conclusion of Sutter et al. who place the first graphene layer at $1.45 \pm 0.1$ Å above the Ru.[12] However, more recent work has indicated a significantly larger *minimum* separation of 2.1 Å between the graphene and the Ru, together with a large corrugation of 1.5 Å in the graphene.[16] This corrugation amplitude is well beyond what we have considered above for Ir(111), and future computations are needed to understand its effect on the low-energy reflectivity.

Let us now turn to the situation for graphene on non-metallic substrates. It is known that on Si-face SiC a graphene-like "buffer layer" exists between the SiC and the first graphene layer. This layer, with 6√3×6√3-R30° surface unit cell, initially forms on the SiC surface in the absence of graphene, although subsequent coverage by graphene is found to largely preserve the structure.[53,54,55,56,57,58,59,60,61,62] The separation between the buffer layer and the graphene layer is observed by surface x-ray reflectivity to be 3.50 Å,[63] somewhat greater than the nominal separation of 3.35 Å between graphene planes, so that an interlayer basis state will certainly exist in that space. However, the buffer layer and the substrate are relatively strongly bonded, with a separation that is only about 2.32 Å.[63] Hence, on the basis of our computation results in Figs. 8 – 13 for other substrates, it is quite plausible that no such state exists between the SiC substrate and the buffer layer (i.e. very large $E_1$ value, referring to Figs. 1 or 10). Therefore, the expected number of reflectivity minima is expected to be $n-1$ where $n$ is the number of graphene layers *including* the buffer layer, in agreement with the experimental result of Hibino et al.[5] and of subsequent authors.[64,65,66,67]

Extending this line of reasoning, for Si-face SiC it is found experimentally that when the buffer layer is transformed to a graphene layer, by hydrogenating or oxidizing the bonds between the buffer layer and the underlying SiC, then an additional reflectivity minimum is formed.[66] We interpret this as a consequence of an interlayer state forming in the space between the substrate and the buffer layer. It appears to be coincidental that the energy $E_1$ at which this state forms is quite close to the energy $E_0$ for the interlayer states that exist in the spaces between the higher graphene layers (our usage of $E_0$ and $E_1$ here is the same as in Figs. 1 and 10). Indeed, with further treatment of the interface, then the new reflectivity minimum shifts considerably with energy, implying an $E_1$ that is different (oftentimes higher) than the $E_0$ value.[68]

Moving to graphene on C-face SiC [the ($000\bar{1}$) surface], for graphene formed in vacuum it is found that, like the Si-face, there are $n$ reflectivity minima for $n$-layers of graphene (i.e. not including the buffer layer in the Si-face count).[69,70] We therefore conclude that there is an open space below the bottom-most graphene layer and whatever layer is below that, with this space having a separation and potential similar to that which exists between the graphene planes themselves (since the observed reflectivity minima form a well-ordered sequence, without an exceptional energy shift of any particular minimum). That is to say, there must be some sort of layer analogous to the Si-face buffer layer that also exists for the C-face formed in vacuum. The detailed structure of any such layer below the graphene is presently unknown, although its unit cell size is reported to be 2×2 and/or 3×3 and it is indeed known to interact only weakly with the bottom-most graphene layer.[59,71] However, for the case of graphene on C-face SiC formed in a neon or a disilane environment, a carbon-rich buffer layer similar to that which forms on the Si-face has been found to form, consisting structurally of a graphene-like layer bonded to the SiC below it.[72] The reflectivity characteristics are then also found to be similar to those of the Si-face, i.e. with a new reflectivity minimum forming when the C-face buffer layer is decoupled from the substrate,[72] consistent with our theoretical interpretation.

## V. Summary

In summary, we have described a computational approach which yields LEER spectra of single- and multilayer graphene on substrates. For $n$ layers graphene with $n \geq 2$, the LEER spectra are found in general to contain $n-1$ distinct minima in the range $0 - 8$ eV, arising from the interlayer states between the pairs of graphene planes. An additional minimum, generally weaker than those just mentioned, can arise from a state localized in the space between the bottom-most graphene plane and the substrate so long as this space is sufficiently wide (i.e. as occurs for weak bonding between that graphene plane and the substrate). By comparing predicted and measured spectra for single-layer graphene we find, for Cu(111), Cu(001), and Ir(111) substrates, separations between the substrate and the graphene of 3.25±0.04, 3.61±0.07 Å, and 3.51±0.05 Å, respectively. For graphene on Pt(111), we find that a structure of a simple weakly-bound graphene layer on the Pt(111), as suggested in Ref. [14], is *not* consistent with the observed LEER spectrum. Rather, a model including intercalated carbon between the Pt surface and the overlying graphene, as proposed in Ref. [11], fits the LEER data better.

The interpretation of LEER spectra described in this work is expected to hold for other 2-dimensional (2D) materials as well, with the interlayer band providing a useful structural probe of the materials (as previously argued by Silkin et al.[9]). Our methodology yields predicted LEER spectra which, by comparison with measured spectra, permits determination of structural characteristics for the material. Our computations use the relatively accurate exchange-correlation term as implemented in VASP, which in the low-energy range make them superior to more traditional LEED I(V) computations[3,11,12,14,49,52] as discussed in Ref. [52]. However, the recent work of Krasovskii et al. provides a method for computing LEER spectra while also providing a good description of the exchange-correlation term. In addition that work includes, at least phenomenologically, an "optical" (imaginary) term in the potential that accounts for electron absorption effects.[26,27] With that term, an improved comparison between experiment and theory is enabled, particularly for higher energies. Another difference between our methodology and that Krasovskii et al. is that we utilize periodic boundary conditions (within VASP), thus necessitating in general multiple computations with different vacuum widths (as in Fig. 8) in order to construct a complete spectrum. However, this requirement is not too onerous for the case of 2D materials, since the dispersion of the interlayer band in those cases leads to a fairly complete LEER spectrum even with just a single computation (as in Figs. 9, 12, and 13).

Further comparing our methodology to that of Krasovskii et al., that work properly includes the full set of complex-wavevector states (i.e. both propagating and evanescent) of the film/substrate system in obtaining the reflectivity. In contrast, our technique, albeit convenient since it makes uses of the readily-available VASP output, makes an approximation with regard to the nature of the states. Since we compare the slab computations to those in a periodically-repeated bulk, we are necessarily only including propagating states of the bulk. On the other hand, our slab computations *do* include evanescent states, for the particular slab thickness that we employ. The evanescent states that we are concerned with here are those that are pseudo-localized at the surface, decaying to some nonzero value in the substrate (slab) and hence potentially contributing in some energy-dependent way to the reflectivity (e.g. see Ref. [73] for a situation with stacking faults in Si). For our computations, the influence on the reflectivity of evanescent waves in the slab can be ascertained by performing computations for various slab thicknesses. For the 2D systems with interlayer states discussed in this work, we have found little dependence on the slab thickness. However, for other systems having more complex electronic structure, it remains to be seen whether a similar independence on slab thickness will be found.

**Acknowledgements**

Discussions with Mark Stiles and Di Xiao are gratefully acknowledged, and we thank S. de la Barrera and P. Mende for their careful reading of the manuscript. This work was supported by the National Science Foundation and by the Office of Naval Research MURI program. The work at Sandia National Laboratories was supported by the US DOE Office of Basic Energy Sciences (BES), Division of Materials Science and Engineering under contract DE-AC04-94AL85000.

**Appendix**

Prior computations have modeled the LEER spectra of graphene using a dynamical theory for low-energy electron diffraction (LEED) intensities.[52] That analysis works well for describing the Bragg peaks in the data (i.e. when an integer multiple of the electron wavelength in the multilayer graphene matches the interplanar spacing). It fails to describe the data in the 0 – 10 eV range, however, since the full electronic structure of the material was not well described by the theory.[52] To describe the minima in the LEER spectra of graphene it is necessary to have a good description of the self-consistent exchange-correlation potential in the material.[6,7] For this purpose we use the Vienna Ab-Initio Simulation Package (VASP), employing the projector-augmented wave method and a generalized-gradient approximation (GGA) for the density functional.[38,39,40,41] We use a plane-wave energy cutoff of 500 eV. We have performed computations using both the standard potentials of VASP as well as potentials designed for GW computations (we do not actually perform GW computations, but we tested these potentials since they are recommended for accuracy of states lying far above the Fermi energy).[74] We obtain essentially identical results for both types of potentials. For graphite, we obtain a band structure identical to that displayed by Hibino et al.[5]

Within VASP (as well as many other electronic structure codes), a supercell is formed that is repeated periodically in all three dimensions. In the $z$-direction, we consider a thin slab of the substrate, on which graphene layers are added on one or both sides. A vacuum region with thickness > 10 Å is then introduced on both sides of this slab. Due to limitations in computation time, the substrate necessarily must be quite thin, whereas of course in the real physical situation the substrate is essentially semi-infinite. In principle, for a thick enough substrate slab, we expect the electronic structure solutions within the center of that slab to approach those of the bulk substrate material. We also make a full electronic structure computation of the bulk substrate, i.e. using no vacuum space in the $z$-direction and with the usual periodic boundary conditions in all directions.

As described in previous work,[6,7] the analysis proceeds in terms of Fourier components of the wavefunction,

$$\psi_{\nu,\mathbf{k}}(\mathbf{r}) = \sum_{G_x,G_y} \phi_{\nu,k_z}^{G_x,G_y}(z)\, e^{i[(k_x+G_x)x+(k_y+G_y)y]} \tag{2}$$

with

$$\phi_{\nu,k_z}^{G_x,G_y}(z) = \sum_{G_z} \frac{C_{\nu,\mathbf{G}}}{\sqrt{V}}\, e^{i(k_z+G_z)z} \tag{3}$$

where $C_{\nu,\mathbf{G}}$ are the plane-wave expansion coefficients for band index $\nu$ and for reciprocal lattice vector of the simulation slab $\mathbf{G} \equiv (G_x, G_y, G_z)$, $V$ is the volume of a unit cell (included for normalization purposes) and $z$ labels the direction normal to the slab. All of our evaluations are performed for $\mathbf{k} = (0,0,k_z)$, i.e. considering an incident plane wave normal to the surface.

Far out in the vacuum, wavefunctions $\psi_{v,k_z}(\mathbf{r})$ of the slab have a specific, separable form: they consist of traveling waves $\exp(i\kappa_{\mathbf{g}} z)$ where $\kappa_{\mathbf{g}}$ labels the $z$-component of the wavevector in the vacuum, multiplied by a sum of lateral waves of the form $A_{\mathbf{g}} \exp[i(g_x x + g_y y)]$ where the lateral wavevector is denoted by $\mathbf{g} = (g_x, g_y)$ and $A_{\mathbf{g}}$ is an amplitude. We have $\kappa_{\mathbf{g}} = \sqrt{2m(E_{v,k} - E_V)/\hbar^2 - g_x^2 - g_y^2}$ where $E_V$ is the vacuum energy (corresponding to the potential energy at a $z$-value sufficiently far from the slab so that the potential is essentially constant; we find that a distance of about 0.5 nm is sufficient for this purpose). The lateral wavevector will correspond to one of the $(G_x, G_y)$ values; $(g_x, g_y) = (0,0)$ for the non-diffracted beam and $(g_x, g_y) \neq (0,0)$ for a diffracted beam, the latter existing only for $E_{v,k} - E_V \geq \hbar^2(g_x^2 + g_y^2)/2m$.

Our approach to solve for the reflectivity from graphene on a thick (semi-infinite) substrate is to first compute the electronic structure for graphene on a thin substrate slab, and then to match those wavefunctions with ones obtained from a solution for the electronic structure in the bulk substrate material. We match to a given state in the bulk having a positive $k_z$ value, that is, corresponding to a wave travelling in the $+z$ direction. The matching is performed for the $(G_x, G_y) = (0,0)$ Fourier component of the wavefunctions, i.e. equating its magnitude and first derivatives between the substrate slab and the substrate bulk. If the substrate slab is sufficiently thick, then its electronic structure should necessarily approach that of the bulk. Thus, even though we are matching only the (0,0) Fourier component we expect, for a thick enough slab, perfect matching for all other components as well. Fortunately, we find that even relatively thin substrate slabs, e.g. containing just 3 or 5 layers of copper, are thick enough to satisfy this criterion. That is, when we match the (0,0) Fourier components, and then investigate the equality between the wavefunctions (and their derivatives) of other Fourier components between the thin slab and the bulk, we find that, indeed, they are very close to equal, so that we have succeeded in matching the full wavefunctions between thin slab and bulk.

To interpret the experimental data, we consider both (111) and (001) orientations of Cu. For our (111) computations we have, unless otherwise specified, assumed a structure in which the graphene layer is strained such that a 1×1 cell of graphene fits onto a surface 1×1 cell of the metal, i.e. for Cu(111) this means a 3.5% expansion of the graphene, as pictured in Fig. 6(b). For the bulk substrate, considering bulk (111)-oriented Cu, we compute the electronic structure of ABC-stacked, i.e. face-centered-cubic (fcc), copper, and we find only a single dispersive NFE band over the energy range 0 – 15 eV relative to the vacuum level, as pictured in Fig. 6(a). For the Cu(001) direction, multiple bands are present in the bulk band structure over that energy range, as seen in Fig. 7(a). However, we find that only a single band has substantial $\phi_{v,k}^{0,0}(z)$ magnitude, i.e., $|\phi_{v,k}^{0,0}(z_0)|\sqrt{V}$ values of order unity with $z = z_0$ corresponding to the location of a plane of Cu atoms, and hence only that band couples to the relevant states of the slab.

We perform the matching between slab and bulk at a $z$ position that corresponds to a copper plane, $z = z_0$. For matching the bulk states to those of graphene on a thin copper slab, we evaluate the logarithmic derivative of the bulk states

$$Q(E, z_0) \equiv -i \left[ \frac{d\phi_{\nu,k_z}^{0,0}(z)/dz}{\phi_{\nu,k_z}^{0,0}(z)} \right]_{z=z_0}, \tag{4}$$

where the quantities in the square brackets are evaluated at $z = z_0$. The dependence of $Q$ on the energy $E$ comes from the $\nu$ and $k_z$ dependence of the bulk wavefunctions. For the dependence of $Q$ on $z_0$, we find for our computations of bulk fcc copper that some such dependence does occur, but evaluating $Q$ at the location of the copper planes we obtain identical values for any of the three planes and furthermore those $Q$ values are purely real. In principle a positive $Q$ value is expected for a wave propagating in the $+z$ direction, although evaluation of Eq. (4) yields both positive and negative values since, in a reduced zone scheme, states associated with negative $G_z$ values (negative $k_z + G_z$) can be folded into the $k_z > 0$ side of the 1st Brillouin zone. In any case, the sign of $Q$ does not affect our evaluation of the reflectivity, below. The inset of Fig. 6(a) shows the $Q$ values we obtain for the (111) direction in Cu.

Returning to the graphene on the thin copper slab, our analysis proceeds similarly to that previously described for free-standing graphene.[6,7] We construct the following two states that are convenient to use in the analysis:

$$\phi_{\nu,+}^{0,0}(z) = [\phi_{\nu,k_z}^{0,0}(z) + \phi_{\nu,-k_z}^{0,0}(z)]/\sqrt{2} \tag{5a}$$

$$\phi_{\nu,-}^{0,0}(z) = [\phi_{\nu,k_z}^{0,0}(z) - \phi_{\nu,-k_z}^{0,0}(z)]/\sqrt{2}. \tag{5b}$$

These states form standing waves, i.e. with real and imaginary parts that are equal within a scale factor (see, e.g., Fig. S1 of the Supplementary Material of Ref. [6]). For convenience, we normalize them such that they are purely real functions. We use these states for two purposes. First, we must ascertain whether or not the particular $\nu, k_z$ state being considered will couple to an incident plane wave in the vacuum. For this purpose we evaluate

$$\sigma_\pm \equiv \frac{\sqrt{A z_S}}{z_S - z_L} \int_{-z_S}^{-z_L} \phi_{\nu,\pm}^{0,0}(z) \exp(i\kappa_0 z) \quad \text{and} \tag{6a}$$

$$\sigma \equiv \left[ |\sigma_+|^2 + |\sigma_-|^2 \right]^{1/2} \tag{6b}$$

where $A$ is the area of the unit cell in the $(x, y)$ plane, $\kappa_0 = \sqrt{2m(E - E_V)}/\hbar$ is the free-electron wavevector where $E_V$ is the vacuum level, $-z_S$ is the far left-hand side of the simulation slab, and where the potential is nearly constant over $-z_S < z < -z_L$. The $\exp(i\kappa_0 z)$ term occurs in the integrand here since we are considering the inner product of $\phi_{v,+}^{0,0}$ and $\phi_{v,-}^{0,0}$ with a plane wave propagating in the $+z$ direction. The values for σ thus obtained have magnitude near unity for the states of interest and are small for other states. However, sometimes "mixed" character of states occurs, as previously described,[6] and for this reason we must use a discriminator for the σ-values that is not much less than 1; a value of 0.8 is used for all results reported in this work and this value works for the present computations to reject all mixed states.

Our second use of the $\phi_{v,+}^{0,0}$ and $\phi_{v,-}^{0,0}$ is to construct a further linear combination such that we obtain a state with a given $Q$ value, at $z = z_0$. The linear combination is written as $\phi_{v,+}^{0,0} + ic\phi_{v,-}^{0,0}$ so that its logarithmic derivative is given by $[(\phi_{v,+}^{0,0})' + ic(\phi_{v,-}^{0,0})']/[\phi_{v,+}^{0,0} + ic\phi_{v,-}^{0,0}]$ where a prime denotes differentiation with respect to $z$. Denoting the location of the central copper plane in the slab by $z_0$, then we equate $-i$ times this logarithmic derivative to the value of $Q$, yielding a value for $c$,

$$c = \left[\frac{Q\phi_{v,+}^{0,0} + i(\phi_{v,+}^{0,0})'}{(\phi_{v,-}^{0,0})' - iQ\phi_{v,-}^{0,0}}\right]_{z=z_0} \tag{7}$$

where the primes again denote differentiation with respect to $z$ and all quantities in the square brackets are evaluated at $z = z_0$. For symmetric potentials this expression simplifies to $c = [Q\phi_{v,+}^{0,0}/(\phi_{v,-}^{0,0})']_{z_0}$ for the case of $z_0 = 0$, from which our construction of the total wavefunction is clear: Around $z = 0$, $\phi_{v,+}^{0,0}$ has the form $A_+ \cos(k_z z)$ and $\phi_{v,-}^{0,0}$ has the form $A_- \sin(k_z z)$ so that $c = QA_+/A_-$, which when inserted into the total wavefunction yields $A_+ \cos(k_z z) + i[QA_+/A_-]A_- \sin(k_z z) = A_+[\cos(k_z z) + iQ\sin(k_z z)]$, thus producing a logarithmic derivative of $iQ$ and hence matching the state from the bulk computation.

With the linear combination $\phi_{v,tot}^{0,0} \equiv \phi_{v,+}^{0,0} + ic\phi_{v,-}^{0,0}$ thus fully determined, we simply evaluate its reflectivity at the far left-hand edge of the simulation slab, $z = -z_S$,

$$R = \left[\left|\frac{i\kappa_0 \phi_{v,tot}^{0,0} - (\phi_{v,tot}^{0,0})'}{i\kappa_0 \phi_{v,tot}^{0,0} + (\phi_{v,tot}^{0,0})'}\right|^2\right]_{z=-z_S} \tag{8}$$

where all quantities in the square brackets are evaluated at $z = -z_S$, and $\kappa_0$ is the free-electron wavevector defined following Eq. (6b). The transmission is given by $T = 1 - R$. This procedure is applicable to substrates that have graphene on the left-hand side or on both sides. If the graphene is placed on the right-hand side, then the evaluations are done at $z = +z_S$ [also, the signs separating the $i\kappa_0\phi$ and $\phi'$ terms in both the numerator and denominator of Eq. (8) are inverted, and the value of $\kappa_0$ is taken as negative, but these sign changes all cancel out].

We note in passing that values for σ can also be computed according to Eqs. (6a) and (6b) but using $\phi_{v,\pm}^{G_x,G_y}(z)$ rather than $\phi_{v,\pm}^{0,0}(z)$ in the integrand of Eq. (6a). Examining those σ-values for nonzero $G_x$ and/or $G_y$ [using a discriminator similar to that just described for the (0,0) beam] permits identification of specific diffracted beams. The intensity of those beams is then obtained by a generalization of our analysis method, a topic that we briefly return to at the end of this Appendix.

Plots of the constructed wavefunctions $\phi_{v,+}^{0,0} + ic\phi_{v,-}^{0,0}$ are shown in Fig. 15, providing results at energies corresponding to the locations of reflectivity minima for graphene-Cu(111) separations of 3.58 and 4.08 Å. The reflectivity minimum at 3.58 Å separation occurs at 4.41 eV; Fig. 15(a) shows the starting state $\phi_{v,k_z}^{0,0}$ at that energy, and Fig. 15(b) shows the final constructed state $\phi_{v,+}^{0,0} + ic\phi_{v,-}^{0,0}$. That final state can be compared with the bulk state shown in Fig. 15(d), and it can be seen that the two wavefunctions are practically identical within the Cu region. Similarly for the reflectivity minimum at 4.08 Å, occurring at 2.16 eV, the wavefunction of the constructed state in the slab, Fig. 15(c), is nearly identical with that in the bulk state shown in Fig. 15(e). For both reflectivity minima, we see that there is a peak in the wavefunction located in the space between the graphene layer and the Cu substrate, indicated by the arrows in Fig. 15. This peak is reminiscent of the wavefunction peaks associated with the interlayer states of free-standing graphene,[6] and hence we attribute the peak to the formation of an interlayer state in the space between the Cu substrate and the bottom-most graphene layer. As this interatomic distance decreases, the wavefunction peak becomes less well defined and the energy of the associated state increases, until the separation reaches about 2.58 Å and the state disappears.

Regarding the validity of our matching process between the states in the thin substrate slab and the bulk material, Fig. 16 shows plots of the wavefunction magnitude and $-i$ times the logarithmic derivative, comparing results from both a 3-layer and 5-layer Cu slab to those from fcc bulk Cu, for the case of 1 layer of graphene on each side of the Cu slab. Results are shown for states in each system located at 4.41 eV above the vacuum level [i.e. corresponding to the states shown in Figs. 15(b) and 15(d)]. For the (0,0) Fourier component, the wavefunction and the logarithmic derivative are matched, i.e. made equal, between the two systems. Examining the values for the other Fourier components, we find that, indeed, they are very nearly equal to each other. Thus, we have essentially achieved matching of all the Fourier components. We do not consider the Cu slab and substrate to be special in this respect, so we tentatively expect a similar level of agreement in the electronic structure between slab and bulk for relatively thin slabs of other substrate materials as well.

Further examining the dependence of the results on the thickness of the slab, we show in Fig. 17 the reflectivity for a bare Cu(111) surface, but now varying the thickness of the Cu slab. For 3 layers the result is the same as the $n = 0$ case of Fig. 8, but for 33 layers a distinct ripple is seen on the computed reflectivity. We interpret the ripple as arising from quantum well states (resonances) in the Cu slab, since its period is in agreement with the expectation for those states.[75] For the 3-layer Cu slab such states have much greater energy spacing between them and hence they do not make a significant impact on the reflectivity (over the energy range shown). We conclude that the quantum well states, while they are artifacts of our methodology, do not represent a significant impediment in the method.

Considering a possible dependence of the results on the strain of the graphene and the registry and/or rotation between graphene and Cu(111), in Table III we list the location of the computed reflectivity minimum $E_1 - E_{VAC}$ for graphene separated by 3.58 Å from Cu(111), for various different matching unit cells between the graphene and the Cu. Results for 1×1 cells with 3.5% expansion of the graphene are listed at the bottom of the table for different registries of the graphene relative to the Cu; the computed $E_1 - E_{VAC}$ values vary by 0.01 eV. This insensitivity to the registry indicates that an incommensurate structure for graphene on Cu(111) (which is known to occur experimentally)[21] should produce $E_1 - E_{VAC}$ within the same range. Examining other results in the table, we find that the $E_1 - E_{VAC}$ values vary by about ±0.05 eV. This relatively small range was somewhat unexpected by us, considering the substantial strains of 10% included there, since for a *free-standing* graphene bilayer we compute that the position of its reflectivity minimum varies by 0.27 eV when a 10% lateral strain is applied. The variation in that case likely arises from the change in the exchange-correlation potential between the two graphene layers due to their changing electron density. For the case of graphene on Cu, it appears that the potential between the graphene and the metal is more determined by the metal itself rather than by the graphene (and/or the influence of a changing work function is different than for the bilayer case). It is also important to note that, in the results of Table III, there is a substantial contribution to the ±0.05 eV range of $E_1 - E_{VAC}$ values arising from the accuracy of the computations. In particular, the entries on the first and third rows, with √3×√3-R30° and 3×3 Cu cells, respectively, have *precisely the same* atomic arrangements; they differ only in the overall size of the cells used for the computation. However, the $E_1 - E_{VAC}$ value for the larger cell is 0.045 eV less than that for the smaller cell (indeed the result for the larger cell is significantly below that for all other results in the table). Hence, our estimated ±0.05 eV range includes the effects of strain, registry, rotation, and computational accuracy.

Moving to the Cu(001) substrate, a surface unit cell is taken to be 1×2, as pictured in Fig. 7(b), with the graphene lattice expanded by 3.5% in the 1× direction and 19.5% in the 2× direction to match. For this relatively large cell, we find that a larger vacuum width in the supercell used for the computations is needed. With 2 nm of vacuum on either side of the graphene-Cu-graphene slab we find that the methodology described above works fine (these results are described below), whereas with only 1 nm of vacuum on either side [as typically used for the (111) orientation] the methodology breaks down because of the predominance of evanescent states localized near the surface. These evanescent states become more predominant as the energies being investigated approach the energies of diffracted beams [i.e. $\hbar^2(g_x^2 + g_y^2)/2m$ as listed following Eq. (3)]; for the (001) unit cell just defined the minimum-energy diffracted beams occur at 5.78 eV for the (0,±1) beams.

The bulk band structure in the (001) direction for the Cu(001)-1×2 unit cell is shown in Fig. 7(a). Because of the larger surface unit cell for this orientation, there are many more bands in this case compared to Fig. 6(a) for Cu(111). However, we find again that only a single band has significant coupling to an incident plane wave; this band is highlighted in red in Fig. 7(a), identified on the basis of $\left|\phi_{\nu,k}^{0,0}(z_0)\right|\sqrt{V} > 0.5$, whereas all other states have $\left|\phi_{\nu,k}^{0,0}(z_0)\right|\sqrt{V} < 10^{-2}$. Note that the onset of the dispersive NFE-like band is *above* the vacuum level for this surface, at 7.15 eV relative to the Fermi level according to theory, or 7.9 eV according to prior experiment, as listed in Table I. Hence the reflectivity will rise to unity for energies below that.

Concerning accuracy of the (001) results, we have considered an alternative unit cell with sides along (3,1) and (−2,2) of the square 1×1 Cu surface unit cell. A 3×3 arrangement of graphene unit cells is placed in that area, with strain of 9.1% in one direction and − 2.4% in the other. The energy of the resulting reflectivity minimum is found to differ from that of the 1×2 computation by 0.15 eV, i.e. a deviation that is larger than the ±0.05 eV range deduced above for the (111) orientation. Allowing for some additional uncertainty in the computational results, and recognizing that the average strain in the graphene for the 1×1 models of graphene on Pt(111) and Ir(111) surfaces does not exceed that of the 1×2 model of graphene on Cu(001), we estimate an overall range of ±0.2 eV for our computational accuracy.

An additional effect that has been considered, for graphene on the Ir(111) surface in particular, is the influence of corrugation in the graphene layer. Figure 18 shows computed reflectivity curves for a flat ($\Delta h = 0$) layer of graphene and for a corrugated 2×2 layers having peak-to-peak corrugation amplitude of $\Delta h = 0.3$ Å. The reflectivity curves acquire a slightly irregular (not smoothly varying) behavior as a result of the corrugation, but nevertheless, it is clear that only a very small shift in the location of the reflectivity minima is produced. These shifts in the location of the minimum (computed by fitting a parabola to the reflectivity, over the energy range 4.5 – 6.8 eV) are summarized in the inset, for both 2×2 and 3×3 corrugations. In the latter case, for corrugation amplitudes as large as 0.6 Å, we find that the resulting shifts are less than ±0.05 eV.

Finally, we briefly consider the situation for diffracted beams. For films with a large lateral unit cell (e.g. as might occur when a commensurate fit of the film material onto a substrate occurs only for some multiple of the primitive unit cell), the resulting lattice constants can be relatively large and diffraction energies small, so that diffracted beams must be explicitly considered even for low incident electron energies. Treating the diffracted beams requires a significant generalization of the theory described above. We have not worked out the details of the general theory, but we provide here a few comments on certain aspects of the procedure. We consider initially a free-standing film.

Diffracted states can be formed when a pair of eigenstates with nonzero $(g_x, g_y)$ components, as described above, are located at the same energy. At this energy these two states will have different $k_z$ values, but nevertheless they have the same $\kappa_\mathbf{g}$ value. For each of these two types of states there are wavefunctions corresponding to $k_z < 0$ and $k_z > 0$, so that we have four states in total to work with. Linear combinations of these four states must be formed in order to satisfy the boundary conditions of an incoming wave from one side of the slab and outgoing non-diffracted waves for both sides as well as a total of four outgoing diffracted waves, from both sides and for the two lateral directions $(g_x, g_y)$ and $(-g_x, -g_y)$. With such combinations, the intensity of the reflected (non-diffracted) wave as well as the intensity of the diffracted waves can be deduced.

For diffracted waves from a film on a substrate, it is necessary to associate the $(g_x, g_y)$ states in the slab computation with bulk states of the same $(g_x, g_y)$ values, again by inspecting their respective $(G_x, G_y)$ Fourier components for similarity. This association between the slab and bulk computational results would appear to be possible for a sufficiently thick slab, although how large that thickness must be is something that remains to be investigated. In any case, given that association between slab and bulk states, then the two would be connected in the same manner as described in Eq. (7). As discussed following Eq. (3), an analysis employing multiple channels (at the same energy) of bulk states connecting to each slab state might be required.

Actually, in principle, diffracted waves should have already been included in our analysis of the Cu(001) surface, since the relatively large unit cell in that case leads to (1,0) and (0,1) diffracted beams for electron energies >5.78 eV. The formation of those beams will lead to a reduction of the intensity of the (0,0) reflected beam, i.e. within a multi-channel analysis. However, we have examined the relevant states that contribute to this diffraction intensity, and we find that, for energies of interest < 9 eV, the (0,0) Fourier components of these states have $\sigma$ values <$10^{-2}$, compared to $\sigma$ values close to unity for the (0,0) components of the non-diffracted states. Thus, inclusion of these diffracted beams will not significantly affect the reflectivities of Fig. 12.

Table I. Summary of experimental results: $E_1 - E_{VAC}$ is the observed energy of the reflectivity minimum for a single layer of graphene on the metal substrate relative to the vacuum level (±0.2 eV) ; $E_{NFE} - E_F$ is the measured energy of the onset of the NFE band; $\Delta\Phi$ is the observed work function difference between the bare metal surface and the graphene-covered surface; $\Phi_M$ is the measured work function of the bare metal surfaces [these surfaces correspond to clean metal surfaces, except for Cu(001) for which it is an oxidized surface]; $\Phi_M - \Delta\Phi$ corresponds to the work function of the graphene-covered surface.

|  | $E_1 - E_{VAC}$ (eV) | $E_{NFE} - E_F$ (eV) | $\Delta\Phi$ (eV) | $\Phi_M$ (eV) | $\Phi_M - \Delta\Phi$ (eV) |
|---|---|---|---|---|---|
| Cu(111) | 7.9[a] | 4.2[d,e] | 0.9[a] | 4.95[h,i] | 4.05 |
| Cu(001) | 4.0[a] | 7.9[d] | 0.8[a] | ≈ 5.0[h,j] | 4.2 |
| Ir(111) | 6.5[b] | 7.6[f] | 0.8[b] | 5.70[k] | 5.10 |
| Pt(111) | none[c] | 6.8[g] | ≈0.2[c] | 5.77[k] | 5.57 |

[a] this work
[b] Ref. [13].
[c] Ref. [14].
[d] Ref. [30].
[e] Ref. [31].
[f] Ref. [32].
[g] Ref. [33].
[h] Ref. [34] and references therein.
[i] Ref. [35].
[j] Ref. [36].
[k] Ref. [37].

Table II. Summary of theoretical results: $a$ is the lattice parameter used in the computations; $E_{NFE} - E_F$ is the computed energy of the onset of the NFE band; $d$ is the value of the graphene-metal separation determined by matching experiment to theory; for that $d$ value, $\Phi_{G-M}$ is the computed value of the work function (vacuum level relative to the Fermi energy) for the graphene-covered metal.

|  | $a$ (Å) | $E_{NFE} - E_F$ (eV) | $d$ (Å) | $\Phi_{G-M}$ (eV) |
|---|---|---|---|---|
| Cu(111) | 3.608[a] | 3.75[b] | 3.25±0.04[b] | 3.95[b] |
| Cu(001) | 3.608[a] | 7.15[b] | 3.61±0.07[b] | 4.61[b] |
| Ir(111) | 3.838[a] | 7.28[b] | 3.51±0.05[b] | 4.96[b] |
| Pt(111) | 3.923[a] | 5.85[b] | interc.[c] | 5.24[b] |

[a] Ref. [42].
[b] this work
[c] LEER experiment of Ref. [14] is inconsistent with a single weakly-bound graphene layer on Pt(111), but it is consistent with a structure including an intercalated carbon layer between the graphene and the Pt(111), as described in Ref. [11].

Table III. Computed energy of the reflectivity minimum, $E_1 - E_{VAC}$, as a function of the commensurate fit between the graphene lattice and the Cu(111) surface lattice, for graphene on a 3-layer Cu slab. The size of the matching cells of the graphene and Cu are listed, along with the resulting strain of the graphene. Registry between the graphene and Cu is shown by listing the location of one graphene atom above a Cu atom, using the usual notation with bulk fcp Cu having ABC stacking on consecutive planes. The Cu stacking in the 3-layer slab is ABA. Listed registry of A, B, or C means at least one carbon atom of the graphene is above that Cu site in the unit cell. For the 1×1 graphene cells, we explicitly show the registries of both carbon atoms in the graphene layer.

| Cu(111) Cell | Graphene Cell | Strain (%) | $E_1 - E_{VAC}$ (eV) | Registry |
|---|---|---|---|---|
| √3×√3-R30° | 2×2 | − 10.3 | 4.439 | A |
| √3×√3-R30° | 2×2 | − 10.3 | 4.442 | B |
| 3×3 | √12×√12-R30° | − 10.3 | 4.394 | A |
| √7×√7-R19.1° | 3×3 | − 8.7 | 4.444 | A |
| √13×√13-R13.9° | 4×4 | − 6.7 | 4.444 | A |
| √21×√21-R10.9° | 5×5 | − 5.1 | 4.445 | A |
| √12×√12-R30° | √13×√13-R13.9° | − 0.5 | 4.476 | A |
| 1×1 | 1×1 | 3.5 | 4.428 | AC |
| 1×1 | 1×1 | 3.5 | 4.430 | AB |
| 1×1 | 1×1 | 3.5 | 4.440 | BC |

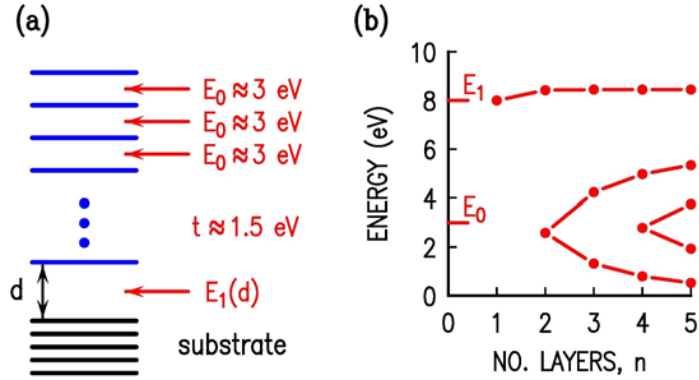

FIG 1. (Color on-line). Schematic view of tight-binding model for determining the energies of reflectivity minima for multilayer graphene on a substrate. (a) Diagram of graphene layers above a substrate, indicating the energies of interlayer states: $E_0$ between graphene planes and $E_1$ between the bottom-most graphene plane and the substrate, with $E_1$ depending on the graphene-substrate separation $d$. The nearest-neighbor coupling between interlayer states is denoted by $t$. (b) Results for a simple tight-binding computation of the energies of the coupled states for $n$ graphene layers, assuming $E_0 = 3$ eV, $E_1 = 8$ eV, and $t = 1.5$ eV.

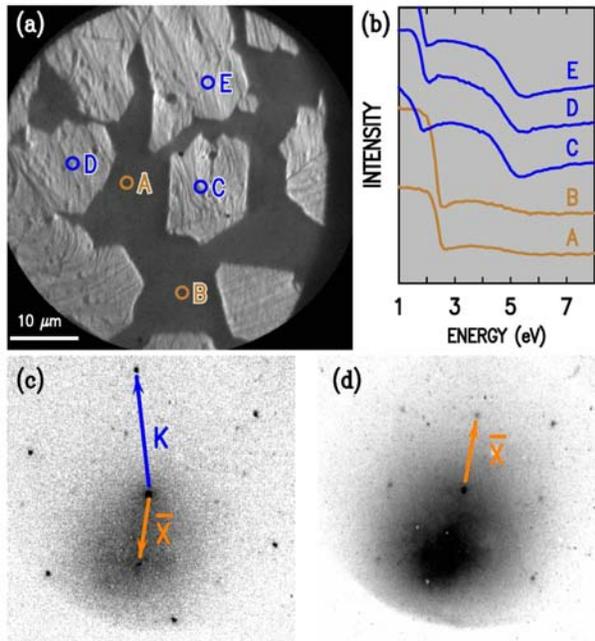

FIG 2. (Color on-line). (a) LEEM image acquired at 4.0 eV, and (b) reflectivity curves from the locations indicated in (a), from sample #1 freshly prepared. (c) and (d) LEED patterns acquired at 44 eV, from bright and dark locations, respectively, nearby those displayed in panel (a). Main (0,1) diffraction spots are labeled in (c) according to the K-point of graphene, and in (d) according to the $\overline{X}$-point of Cu(001). The sample consists of graphene on a Cu foil, prepared by CVD and transferred through air to the LEEM.

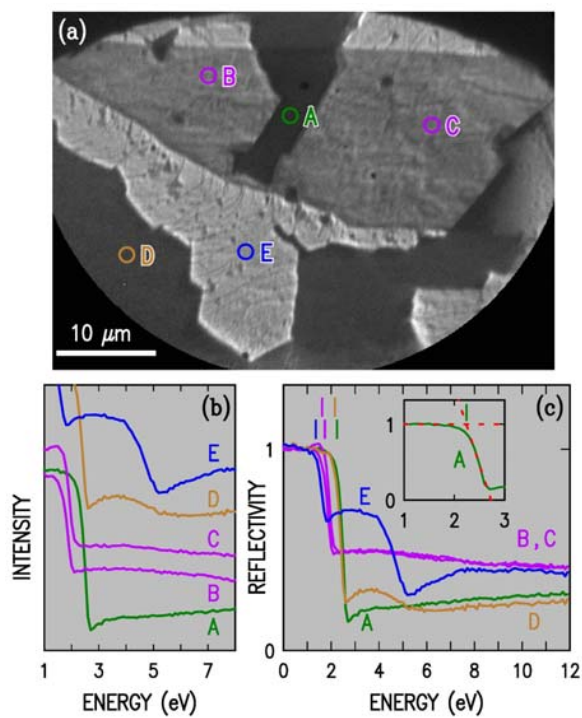

FIG 3. (Color on-line). (a) LEEM image acquired at 2.8 eV, (b) and (c) reflectivity curves, from the same samples as Fig. 1, freshly prepared. The inset of (c) illustrates the method for determining the energy at which the reflectivity rises to unity, using reflectivity curve A.

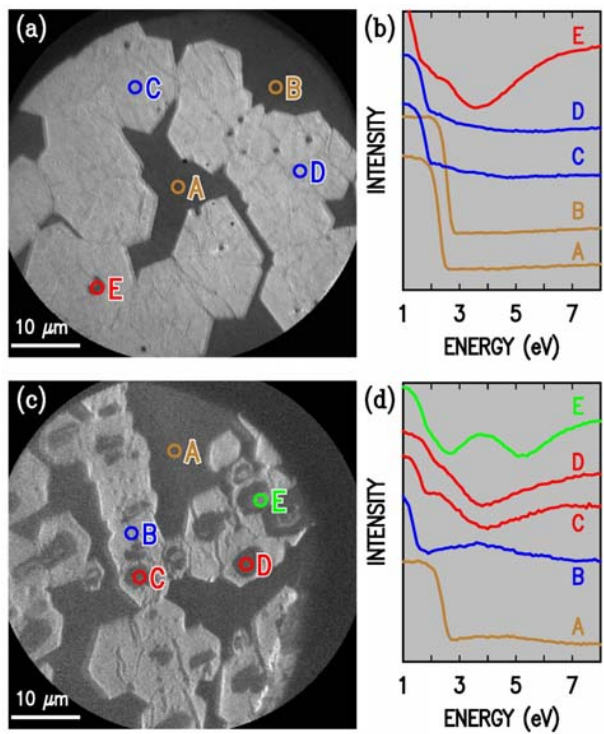

FIG 4. (Color on-line). (a) LEEM image acquired at 4.0 eV from sample #1 after air exposure for several months, and (b) reflectivity curves from the locations indicated in (a). (c) LEEM image acquired at 4.0 eV from sample #2 after air exposure for several months, and (d) reflectivity curves from the locations indicated in (c). The samples consist of graphene on a Cu foil, prepared by CVD and transferred through air to the LEEM.

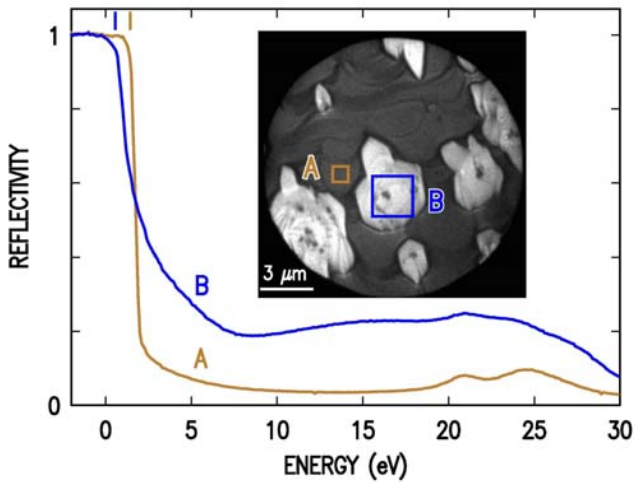

FIG 5. (Color on-line) Reflectivity curves acquired from sample #3, graphene on a Cu(111) single crystal, from the locations indicated in the LEEM image displayed in the inset (image acquired at 3.0 eV). Region B has a single graphene layer while region A is bare Cu. The sample was prepared and characterized in the LEEM, all under ultra-high vacuum conditions.

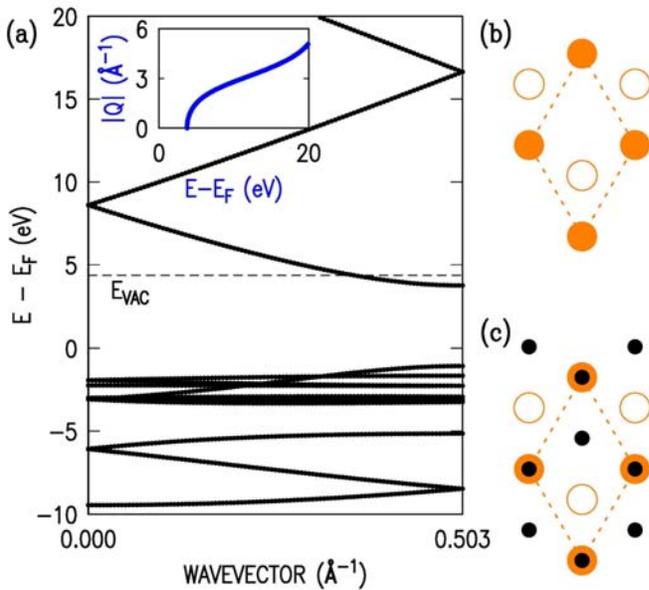

FIG 6. (Color on-line). (a) Band structure for bulk fcc Cu with wavevector varying in the (111) direction. The vacuum level ($E_{VAC}$) is indicated, for a graphene layer located 3.58 Å above a Cu(111) surface. A dispersive band is evident, starting at 0.6 eV below $E_{VAC}$. The inset shows the logarithmic derivative of the wavefunction for the states of that band, evaluated on a plane of Cu atoms. (b) Schematic model for the Cu(111) bulk unit cell employed in the band structure computation (1st layer Cu atoms shown as filled orange circles, 2nd layer Cu atoms as open orange circles, 3rd layer Cu atoms not shown). The lateral part of the unit cell is indicated by dashed lines. (c) Schematic model for graphene on a Cu(111) slab (carbon atoms shown in black).

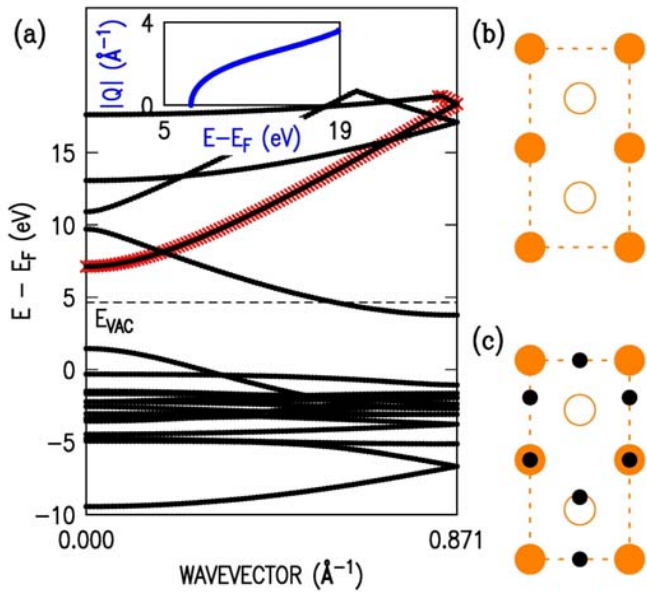

FIG 7. (Color on-line). (a) Band structure for bulk fcc Cu with wavevector varying in the (001) direction. The vacuum level ($E_{VAC}$) is indicated, for a graphene layer located 3.58 Å above a Cu(001) surface. A dispersive band is apparent, beginning to 2.5 eV above $E_{VAC}$, as highlighted with red x-marks. The inset shows the logarithmic derivative of the wavefunction for the dispersive states, evaluated on a plane of Cu atoms. (b) Schematic model for the Cu(001) bulk unit cell employed in the band structure computation (1st Cu layer atoms shown as filled orange circles, 2nd Cu layer atoms as open orange circles, 3rd Cu layer atoms not shown). The lateral part of the unit cell is indicated by dashed lines. (c) Schematic model for graphene on a Cu(001) slab (carbon atoms shown in black).

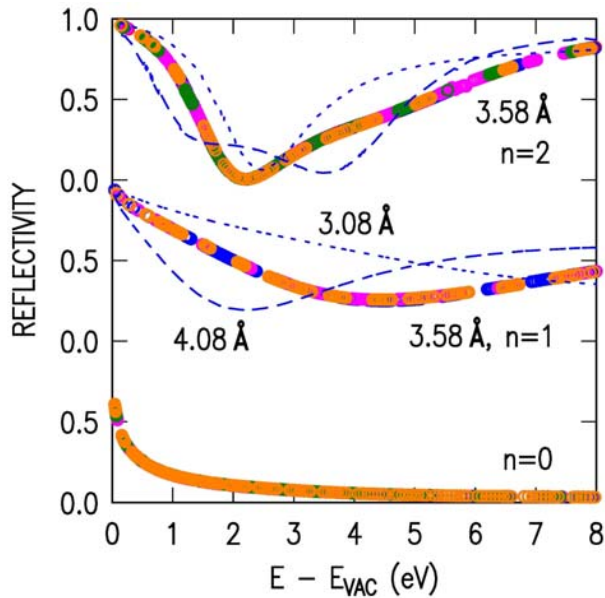

FIG 8. (Color on-line). Computed reflectivity for various thicknesses $n$ of multilayer graphene on a Cu(111) surface. Computations are for a nominal separation between the Cu and the graphene of 3.58 Å (solid symbols), with results also shown for separations of 4.08 Å (dashed lines) and 3.08 Å (dotted lines). For each $n$, a series of computations are performed with different vacuum widths; differently shaded (colored) data points are used for plotting the results for each width.

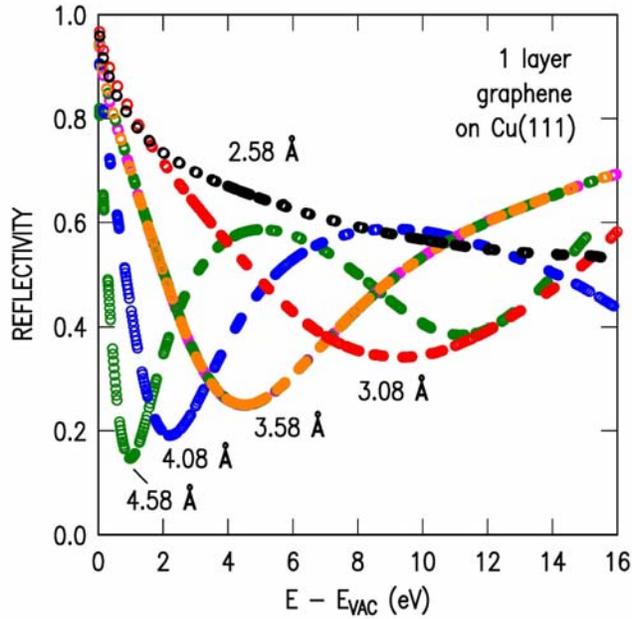

FIG 9. (Color on-line) Computed reflectivity for 1 layer of graphene on a Cu(111) surface. Different curves correspond to different separations between the graphene and the Cu surface, as listed.

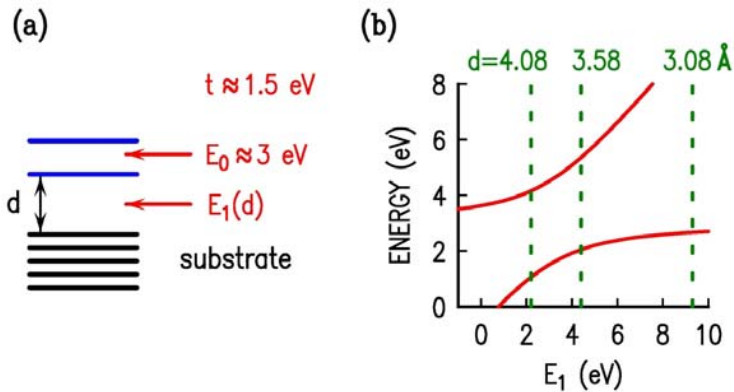

FIG 10. (Color on-line). Schematic view of tight-binding model for determining the energies of reflectivity minima for $n = 2$ layers graphene on a substrate. (a) Diagram of graphene layers above a substrate, indicating the energies of interlayer states: $E_0$ between graphene planes and $E_1$ between the bottom-most graphene plane and the substrate, with $E_1$ depending on the graphene-substrate separation $d$. The nearest-neighbor coupling between interlayer states is denoted by $t$. (b) Results for a simple tight-binding computation of the energies of the coupled states, as a function of the $E_1$ energy, assuming $E_0 = 3\,\text{eV}$ and $t = 1.5\,\text{eV}$. Dotted lines show the correspondence with the first-principles computations of the reflectivity spectra for 2 layers of graphene on Cu(111), from Fig. 8.

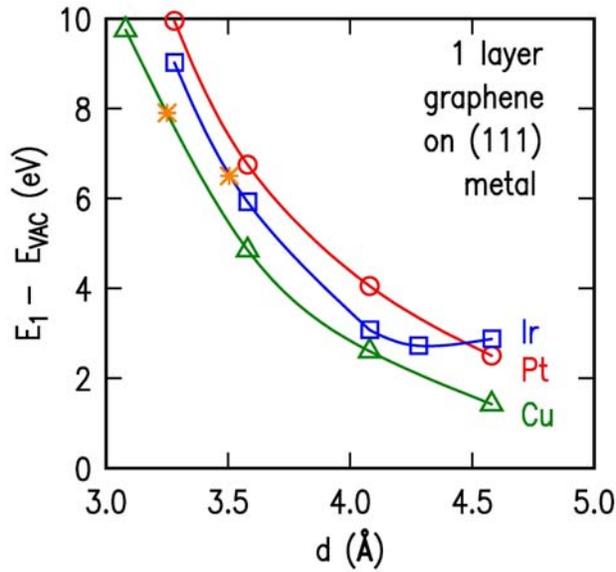

FIG 11. (Color on-line). Summary of the theoretical energies $E_1 - E_{VAC}$ of the reflectivity minima for one graphene layer on a (111) metal substrate, for the metals listed, as a function of the graphene-metal separation. The energies have been shifted by the systematic error in the computations, as summarized by Eq. (1) of the text. Star symbols (orange, in the color version) show the experimental values for the reflectivity minima for Cu and Ir, with no such minimum observed for Pt.

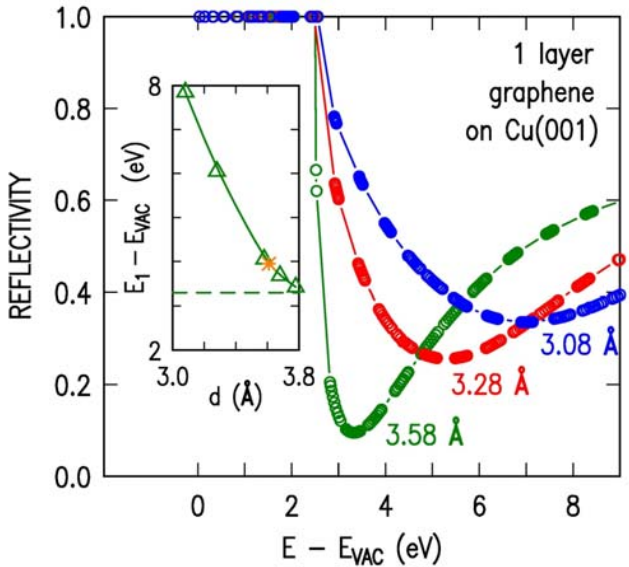

FIG 12. (Color on-line). Computed reflectivity for 1 layer of graphene on a Cu(001) surface. Different curves correspond to different separations between the graphene and the Cu surface, as listed. Inset: theoretical locations of the minima in the computed reflectivity, shifted by the systematic error $\delta E_{NFE} = 0.75$ eV in the computations as summarized by Eq. (1) in the text. The dashed line shows the onset of the NFE band, and the star symbol (orange, in the color version) shows the *observed* energy of the reflectivity minimum. (The $\delta E_{NFE}$ correction has *not* been applied to the reflectivity curves in the main part of the figure).

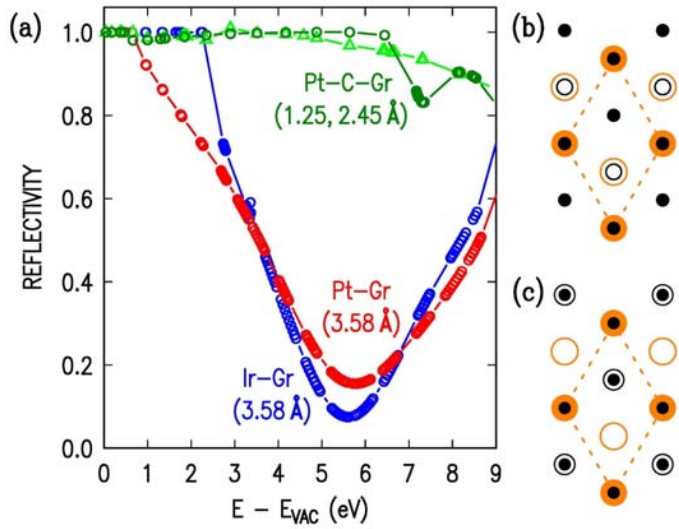

FIG 13. (Color on-line). (a) Computed reflectivity for graphene on Pt(111) and Ir(111) substrates. Results are provided for a graphene layer on the bare metal substrate, Pt-Gr and Ir-Gr, with 3.58 Å spacing between the metal and the graphene, and for Pt-C-Gr structures with an intercalated carbon layer between a Pt substrate and a graphene layer (1.25 Å spacing between the Pt and the intercalated C layer and 2.45 Å between the intercalant layer and the graphene). (b) and (c) Top views of the Pt-C-Gr structures, with first carbon layer atoms shown as filled black circles, second carbon layer atoms as open black circles, first Pt layer atoms as filled orange circles, and second Pt layer atoms as open orange circles. The reflectivity for Pt-C-Gr model (b) is shown by the green circles in (a), and for Pt-C-Gr model (c) by the green triangles in (a).

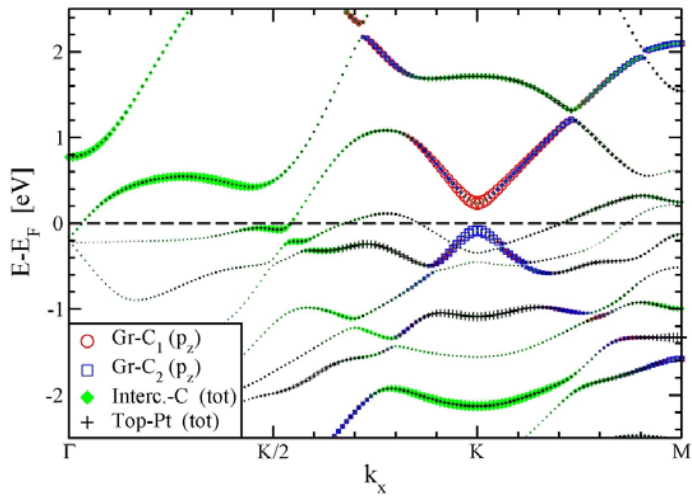

FIG 14. (Color on-line) Band structure with wavevector varying in lateral direction, for the Pt-C-Gr model with an intercalated carbon layer between the graphene and the Pt(111) surface. Different symbols denote the wavefunction character on the two carbon atoms of the graphene layer ($C_1$ and $C_2$), the intercalated carbon layer (Interc.-C), and the top Pt layer. The size of the symbols corresponds to the respective wavefunction magnitude.

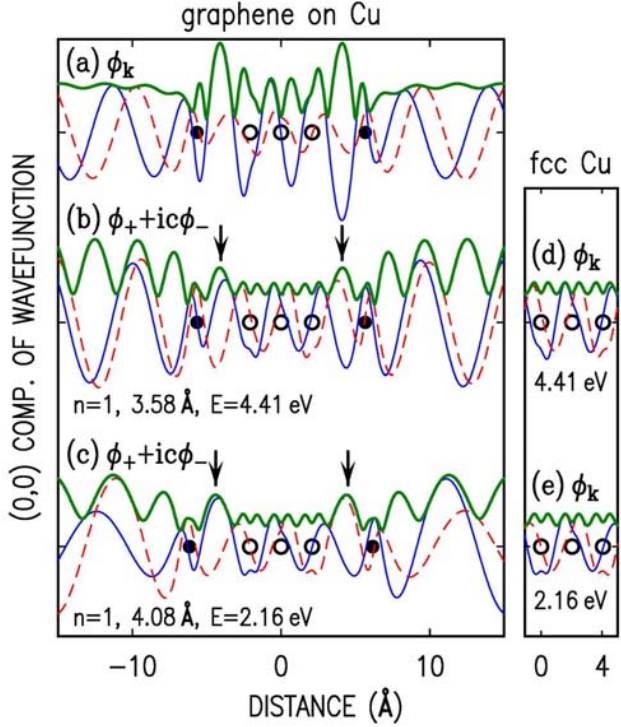

FIG 15. (Color on-line). (0,0) components of the wave-function for graphene on a 3-layer Cu slab (left) and for bulk fcc Cu (right). For the graphene on Cu, results are shown for two different separations and at energies (relative to $E_{VAC}$) corresponding to the reflectivity minimum: 3.58 Å and 4.41 eV for (a) and (b), and 4.08 Å and 2.16 eV for (c). Plot (a) shows the wavefunction at a particular wavevector, $\phi_{k_z}^{0,0}$, with (b) and (c) showing the constructed wavefunction $\phi_{V,+}^{0,0} + ic\phi_{V,-}^{0,0}$ (see text). For the bulk Cu, plots (d) and (e) show the wavefunction at a particular wavevector, $\phi_{k_z}^{0,0}$, at the energies listed [matched to the energies of the states in (b) and (c), respectively]. The real part of the wavefunction is shown by the thin solid line, the imaginary part by the thin dashed line, and the magnitude by the thick solid line (blue, red, and green, respectively, in the color version). The solid black dots indicate the positions of the graphene layers and the open circles indicate Cu layers. The arrows indicate peaks in the wavefunctions that are concentrated between the graphene layers and the Cu substrate.

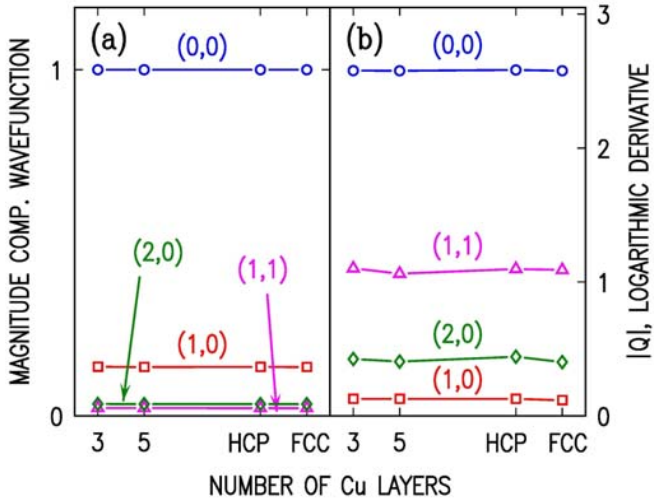

FIG 16. (Color on-line). (a) Magnitude of various Fourier components of the wavefunctions, and (b) the logarithmic derivative, as a function of the number of copper layers in the slab (3 or 5) compared to bulk (both hcp and fcc) computations. For the (0,0) component, both the wavefunction and the logarithmic derivative are matched between slab and bulk, and for the other components the level of agreement for these quantities between slab and bulk (i.e. the near constancy with respect to the number of layers) is a test of the validity of the theoretical method.

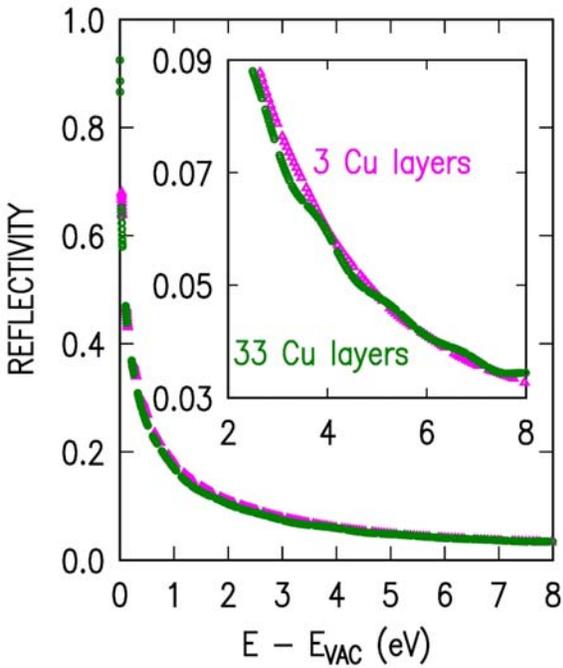

FIG 17. (Color on-line). Computed reflectivity for slabs of Cu(111), with slab thickness of 3 layers (triangle symbols) and 33 layers (circle symbols).

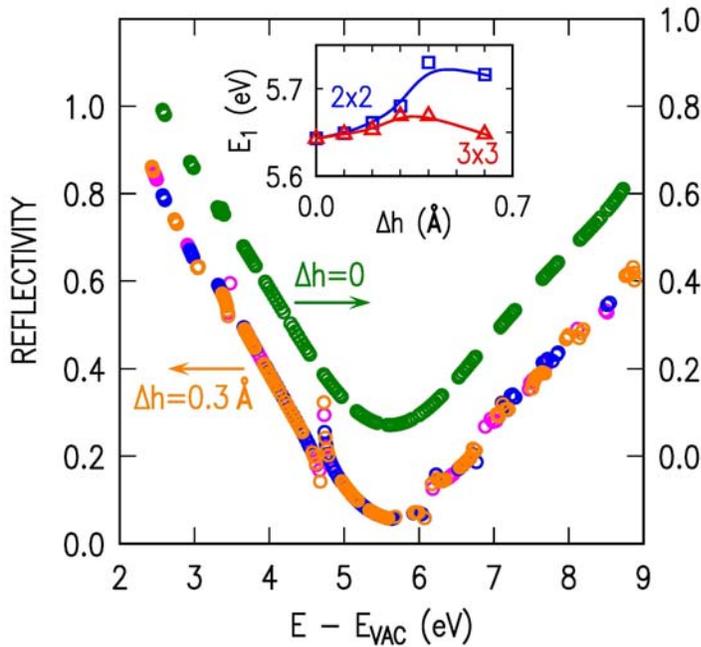

FIG 18. (Color on-line). Computed reflectivity for graphene on Ir(111), assuming peak-to-peak corrugation amplitude $\Delta h$ of the graphene. Main plot shows reflectivity curves for $\Delta h$ values of 0 and 0.3 Å, for a 2×2 corrugation period. Inset shows the resulting shifts in the location of the reflectivity minimum, for both 2×2 and 3×3 corrugation periods.

# References

[1] B. T. Jonker, N. C. Bartelt, and R. L. Park, Surf. Sci. **127**, 183 (1983).
[2] R. Zdyb and E. Bauer, Phys. Rev. Lett. **88**, 166403 (2002).
[3] W. F. Chung, Y. J. Feng, H. C. Poon, C. T. Chan, S. Y. Tong, and M. S. Altman, Phys. Rev. Lett. **90**, 216105 (2003).
[4] M. Altman, J. Phys.: Condens. Matter **17**, S1305 (2005).
[5] H. Hibino, H. Kageshima, F. Maeda, M. Nagase, Y. Kobayashi, and H. Yamaguchi, Phys. Rev. B **77**, 075413 (2008).
[6] R. M. Feenstra, N. Srivastava, Q. Gao, M. Widom, B. Diaconescu, T. Ohta, G. L. Kellogg, J. T. Robinson, and I. V. Vlassiouk, Phys. Rev. B **87**, 041406(R) (2012).
[7] R. M. Feenstra and M. Widom, to appear in Ultramicroscopy, arXiv:1212:5506 (2012).
[8] M. Posternak, A. Baldereschi, A. J. Freeman, E. Wimmer, and M. Weinert, Phys. Rev. Lett. **50**, 761 (1983).
[9] V. M. Silkin, J. Zhao, F. Guinea, E. V. Chulkov, P. M. Echenique, and H. Petek, Phys. Rev. B **80**, 121408(R) (2009), and references therein.
[10] D. A. Stewart, Computing in Science and Engineering **14**, 55 (2012).
[11] H. Zi-pu, D. F. Ogletree, M. A. Van Hove, and G. A. Somorjai, Surf. Sci. **180**, 433 (1987).
[12] P. W. Sutter, J.-I. Flege, and E. A. Sutter, Nature Mat. **7**, 406 (2008).
[13] E. Loginova, S. Nie, K. Thürmer, N. C. Bartelt, and K. F. McCarty, Phys. Rev. B **80**, 085430 (2009).
[14] P. Sutter, J. T. Sadowski, and E. Sutter, Phys. Rev. B **80**, 245411 (2009).
[15] G. Otero, et al., Phys. Rev. Lett. **105**, 216102 (2010).
[16] W. Moritz, B. Wang, M.-L. Bocquet, T. Brugger, T. Greber, J. Wintterlin, and S. Günther, Phys. Rev. Lett. **104**, 136102 (2010).
[17] C. Busse, et al., Phys. Rev. Lett. **107**, 036101 (2011).
[18] S. Nie, J. M. Wofford, N. C. Bartelt, O. D. Dubon, and K. F. McCarty, Phys. Rev. B **84**, 155425 (2011).
[19] Z. Sun, S. K. Hämäläinen, J. Sainio, J. Lahtinen, D. Vanmaekelbergh, and P. Liljeroth, Phys. Rev. B **83**, 081415 (2011).
[20] S. Nie, W. Wu, S. Xing, Q. Yu, J. Bao, S.-S. Pei, and K. F. McCarty, New J. Phys. **14**, 093028 (2012).
[21] Z. R. Robinson, P. Tyagi, T. R. Mowll, C. R. Ventrice, and J. B. Hannon, Phys. Rev. B **86**, 235413 (2012).
[22] A. Dahal, R. Addou, P. Sutter, and M. Batzill, Appl. Phys. Lett. **100**, 241602 (2012).
[23] X. S. Li, W. W. Cai, J. H. An, S. Kim, J. Nah, D. X. Yang, R. Piner, A. Velamakanni, I. Jung, E. Tutuc, S. K. Banerjee, L. Colombo, and R. S. Ruoff, Science **324**, 1312 (2009).
[24] M. Vanin, J. J. Mortenson, A. K. Kelkkanen, J. M. Garcia-Lastra, K. S. Thygesen, and K. W. Jacobsen, Phys. Rev. B **81**, 081408(R) (2010).
[25] T. Olsen, J. Yan, J. J. Mortensen, and K. S. Thygesen, Phys. Rev. Lett. **107**, 156401 (2011).
[26] E. E. Krasovskii, W. Schattke, V. N. Strocov, and R. Claessen, Phys. Rev. B **66**, 235403 (2002).
[27] E. E. Krasovskii and V. N. Strocov, J. Phys.: Condens. Matter **21**, 314009 (2009).
[28] I. Vlassiouk, M. Regmi, P. Fulvio, S. Dai, P. Datskos, G. Eres, and S. Smirnov, ACS Nano, **5**, 6069 (2011).
[29] C. M. Yim, K. L. Man, X. Xiao, and M. S. Altman, Phys. Rev. B **78**, 155439 (2008).

[30] J. A. Knapp, F. J. Himpsel, and D. E. Eastman, Phys. Rev. B **19**, 4952 (1979).
[31] J. F. Janak, A. R. Williams, and V. L. Moruzzi, Phys. Rev. B **11**, 1522 (1975).
[32] J. F van der Veen, F. J. Himpsel, and D. E. Eastman, Phys. Rev. B **22**, 4226 (1980).
[33] H. Wern, R. Courths, G. Leschik, and S. Hüfner, Z. Phys. B – Condensed Matter **60**, 292 (1985).
[34] N. V. Smith, Phys. Rev. B **32**, 3549 (1985).
[35] K. Takeuchi, A. Suda, and S. Ushioda, Surf. Sci. **489**, 100 (2001).
[36] A. V. Ermakov, E. Z. Ciftlikli, S. E. Syssoev, I. G. Shuttleworth, and B. J. Hinch, Rev. Sci. Instru. **81**, 105109 (2012).
[37] M. Kaack and D. Fick, Surf. Sci. **342**, 111 (1995).
[38] G. Kresse and J. Hafner, Phys. Rev. B **47**, RC558 (1993).
[39] G. Kresse and J. Furthmuller, Phys. Rev. B **54**, 11169 (1996).
[40] G. Kresse, and D. Joubert, Phys. Rev. B **59**, 1758 (1999).
[41] J. P. Perdew, K. Burke, and M. Ernzerhof, Phys. Rev. Lett. **77**, 3865 (1996); *ibid*., Erratum, Phys. Rev. Lett. **78**, 1396 (1997).
[42] P. Villars and L. D. Calvert, *Pearson's Handbook of Crystallographic Data for Intermetallic Phases* (American Society for Materials, Materials Park, Ohio, 1991), 2nd edition.
[43] N. V. Smith, Appl. Surf. Sci. **22/23**, 349 (1985).

[44] E.g., F. Tran, Phys. Lett. A **376**, 879 (2012).
[45] I. G. Kim, B. C. Lee, and J. I. Lee, Surf. Rev. Lett. 10, **207** (2003).
[46] J. L. F. Da Silva, C. Stampfl, and M. Scheffler, Surf. Sci. 600, 703 (2006).
[47] To determine the precise location of the onset of the NFE band from a LEER spectrum, the influence of the optical potential must be considered, as illustrated e.g. in Fig. 2 of Ref. [26]. Hence, for our graphene-on-Cu(001) spectra such as curve E of Fig. 3, the onset of the NFE band likely occurs a few tenths of an eV *above* 4.2 eV, thereby producing a value for the onset relative to the vacuum level that is >2.9 eV (and thus producing better agreement with the expected value of $7.9 - 4.61 = 3.29$ eV).
[48] G. Csányi, P. B. Littlewood, A. H. Nevidomskyy, C. J. Pickard, and B. D. Simons, Nature Phys. **1**, 42 (2005).
[49] M. A. Van Hove, W. Moritz, H. Over, P. J. Rous, A. Wander, A. Barbieri, N. Materer, U. Starke, and G. A. Somorjai, Surf. Sci. Rep. **19**, 191 (1993).
[50] D. Rolles, R. Díez Muiño, F. J. García de Abajo, C. S. Fadley, and M. A. Van Hove, J. Electron Spectrosc. Related Phenom. **114-116**, 107 (2001).
[51] V. Blum and K. Heinz, Comput. Phys. Commun. **134**, 392 (2001).
[52] H. Hibino, S. Mizuno, H. Kageshima, M. Nagase, and H. Yamaguchi, Phys. Rev. B **80**, 085406 (2009).
[53] F. Owman and P. Mårtensson, Surf. Sci. **369**, 126 (1996).
[54] I. Forbeaux, J.-M. Themlin, and J.-M. Debever, Phys. Rev. B **58**, 16396 (1998).
[55] W. Chen, H. Xu, L. Liu, X. Gao, D. Qi, G. Peng, S. C. Tan, Y. Feng, K. P. Loh, and A. T. S. Wee, Surf. Sci. **596**, 176 (2005).
[56] Th. Seyller, K. V. Emtsev, K. Gao, F. Speck, L. Ley, A. Tadich, L. Broekman, J. D. Riley, R. C. G. Leckey, O. Rader, A. Varykhalov, and A. M. Shikin, Surf. Sci. **600**, 3906 (2006).
[57] C. Riedl, U. Starke, J. Bernhardt, M. Franke, and K. Heinz, Phys. Rev. B **76**, 245406 (2007).
[58] F. Varchon, R. Feng, J. Hass, X. Li, B. Ngoc Nguygen, C. Naud, P. Mallet, J.-Y. Veuillen, C. Berger, E. H. Conrad, and L. Magaud, Phys. Rev. Lett. **99**, 126805 (2007).

[59] K. V. Emtsev, F. Speck, Th. Seyller, L. Ley, and J. D. Riley, Phys. Rev. B **77**, 155303 (2008).
[60] S. W. Poon, and W. Chen, E. S. Tok, and A. T. S. Wee, Appl. Phys. Lett. **92**, 104102 (2008).
[61] J. B. Hannon and R. M. Tromp, Phys. Rev. B 77, 241404 (2008).
[62] Y. Qi, S. H. Rhim, G. F. Sun, M. Weinert, and L. Li, Phys. Rev. Lett. **105**, 085502 (2010).
[63] J. Hass, J. E. Millan-Otoya, P. N. First, and E. H. Conrad, Phys. Rev. B **78**, 205424 (2008).
[64] T. Ohta, F. El Gabaly, A. Bostwick, J. L. McChesney, K. V. Emtsev, A. K. Schmid, T. Seyller, K. Horn, and E. Rotenberg, New Jouranl of Physics 10, 023034 (2008).
[65] C. Virojanadara, M. Syväjarvi, R. Yakimova, L. I. Johansson, A. A. Zakharov, and T. Balasubramanian, Phys. Rev. B **78**, 245403 (2008).
[66] C. Riedl, C. Coletti, T. Iwasaki, A. A. Zakharov, and U. Starke, Phys. Rev. Lett. **103**, 246804 (2009).
[67] Luxmi, N. Srivastava, R. M. Feenstra, and P. J. Fisher, J. Vac. Sci. Technol. **28**, C5C1 (2010).
[68] K. V. Emtsev, A. A. Zakharov, C. Coletti, S. Forti, and U. Starke, Phys. Rev. B **84**, 125423 (2011).
[69] H. Hibino, H. Kageshima, F. Maeda, M. Nagase, Y. Kobayashi, Y. Kobayashi, and H. Yamaguchi, E-J. Surf. Sci. Nanotech. **6**, 107 (2008).
[70] Luxmi, N. Srivastava, G. He, R. M. Feenstra, and P. J. Fisher, Phys. Rev. B **82**, 235406 (2010).
[71] F. Hiebel, P. Mallet, F. Varchon, L. Magaud, and J.-Y. Veuillen, Phys. Rev. B **78**, 153412 (2008).
[72] N. Srivastava, G. He, Luxmi, and R. M. Feenstra, Phys. Rev. B **85**, 041404(R) (2012).
[73] M. Stiles and D. R. Hamann, Phys. Rev. B **41**, 5280 (1990).
[74] http://www.vasp.at/index.php/news/36-highlights/100-new-release-paw-datasets
[75] D. J. Griffiths, Introduction of Quantum Mechanics (Prentice Hall, Upper Saddle River, New Jersey, 1995), p.64.